\documentclass[useASM]{mn2e}
\usepackage{graphicx}
\begin{document}

\title{Towards the Properties of Long Gamma-Ray Burst Progenitors with
{\em Swift} Data\footnote{Send offprint request to: Enwei Liang
(lew@gxu.edu.cn)}}

\author[Cui et al. ]
       {Xiao-Hong Cui$^{1}$, En-Wei Liang$^{2}$, Hou-Jun Lv$^{2}$, Bin-Bin Zhang$^{3}$, and Ren-Xin Xu$^{1}$
 \\
 $^1$%Department of Astronomy, Peking University, Beijing 100871, China\\
 School of Physics and State Key Laboratory of Nuclear Physics and Technology, Peking University, Beijing 100871, China\\
 $^2$Department of Physics, Guangxi University, Nanning 530004, China\\
$^{3}$Department of Physics and Astronomy, University of Nevada, Las Vegas, NV
89154 } \maketitle

\label{firstpage}
\begin{abstract}
We investigate the properties of both the prompt and X-ray
afterglows of gamma-ray bursts (GRBs) in the burst frame with  a
sample of 33 {\em Swift} GRBs. Assuming that the steep decay segment
in the canonical X-ray afterglow lightcurves is due to the curvature
effect, we fit the lightcurves with a broken power-law to derive the
zero time of the last emission epoch of the prompt emission ($t_1$)
and the beginning as well as the end time of the shallow decay
segment ($t_2$ and $t_3$). We show that both the isotropic peak
gamma-ray luminosity ($L_{\rm peak, \gamma}$) and gamma-ray energy
($E_{\rm iso, \gamma}$) are correlated with the isotropic X-ray
energy ($E_{\rm iso, X}$) of the shallow decay phase and the
isotropic X-ray luminosity at $t_2$ ($L_{\rm X, t_{2}}$). We infer
the properties of the progenitor stars based on a model proposed by
Kumar et al. who suggested that both the prompt gamma-rays and the
X-ray afterglows are due to the accretions of different layers of
materials of the GRB progenitor star by a central black hole (BH).
We find that most of the derived masses of the core layers are
$M_c=0.1\sim 5 M_\odot$, and their average accretion rates in the
prompt gamma-ray phase are $\dot{M_c}=0.01\sim 1 M_\odot$/s, with a
radius of $r_c=10^8\sim 10^{10}$ cm. The rotation parameter is
correlated with the burst duration, being consistent with the
expectation of collapsar models. The estimated radii and the masses
of the fall-back materials for the envelope layers are
$r_e=10^{10}\sim 10^{12}$ cm and $M_e=10^{-3}\sim 1M_\odot$,
respectively. The average accretion rates in the shallow decay phase
are correlated with those in the prompt gamma-ray phase, but they
are much lower, i.e., $\dot{M}_e=10^{-8}\sim 10^{-4}M_\odot$/s. The
$r_e$ values are smaller than the photospheric radii of Wolf-Rayet
(WR) stars. In our calculation, we assume a uniform mass of the
central BH ($M_{\rm BH}=10M_\odot$). Therefore, we may compare our
results with simulation results. It is interesting that the
assembled mass density profile for the bursts in our sample is well
consistent with the simulation for a pre-supernova star with mass
$M=25M_\odot$.
\end{abstract}

\section {Introduction}
One of the unexpected findings with the X-Ray Telescope (XRT)
on-board the gamma-ray burst (GRB) mission {\em Swift} is the
discovery of a canonical X-ray lightcurve, which shows successively
four power-law decay segments with superimposed erratic flares
(Zhang et al. 2006; Nousek et al. 2006). It starts with an initial
steep decay following the prompt emission. This phase usually lasts
hundreds of seconds and could be generally explained as the tail
emission of the prompt GRB due to the curvature effect (Kumar \&
Panaitescu 2000; Qin et al. 2004; Qin 2008; Zhang et al. 2006; Liang
et al. 2006; Zhang et al. 2007a, 2009). A shallow decay segment,
which lasts from hundreds to thousands of seconds, is usually seen
following the GRB tail (O'Brien et al. 2006; Liang et al. 2007). It
transits to a normal decay segment or a sharp drop (Troja et al.
2007; Liang et al. 2007).

Phenomenologically, the canonical lightcurves are well fitted with a
two-component model (Willingale et al. 2007; Ghisellini 2008), but
the physics that shapes the canonical XRT lightcurves is unclear
(Zhang 2007). The origin of the shallow decay segment is under
debate. The normal decay segments following the shallow decay
segment are roughly consistent with the forward shock models
(Willingale et al. 2007; Liang et al. 2007), favoring the long
lasting energy injection models for the shallow-decay segments
(Zhang 2007). The chromatic transition time observed in both the
X-ray and optical afterglows challenges this scenario (Fan \& Piran
2006; Panaitescu 2007; Liang et al. 2007).

Alternative models were proposed to explain the shallow decay
segment(see review by Zhang 2007). The shallow decay would result in
a high gamma-ray efficiency (e.g. Zhang et al. 2007), and Ioka et
al. (2006) proposed that the efficiency crisis may be avoided if a
weak relativistic explosion occurs $10^3-10^6$ s prior to the main
burst or if the microphysical parameter of the electron energy
increases during the shallow decay. Shao \& Dai (2007) interpreted
the X-ray lightcurve as due to dust scattering of some prompt
X-rays(c.f., Shen et al. 2009). The scattering of the external
forward-shock or of the internal shock synchrotron emission by a
relativistic outflow could also explain the observed X-ray
afterglows (Shen et al. 2006; Panaitscu 2007). Uhm \& Beloborodov
(2007) and Genet, Daigne \& Mochkovitch (2007) interpreted both
X-ray and optical afterglow as emission from a long-lived reverse
shock. Liang et al. (2007) argued that the physical origin of the
shallow decay segment may be diverse and those shallow decay
segments following an abrupt cutoff might be of internal origin (see
also Troja et al. 2007). Ghisellini et al. (2007) suggested that the
shallow-to-normal transition in the X-ray afterglows may be produced
by late internal shocks, and the transition is due to the jet effect
in the prompt ejecta (see also Nava et al. 2007). Racusin et al.
(2009) also suggested that the shallow-to-normal transition may be a
jet break occurring during energy injection. Interestingly, Yamazaki
(2009) recently suggested that the X-ray emission might be an
independent component prior to the GRB trigger. By shifting the zero
time point of the shallow-to-normal decay segment in the canonical
XRT lightcurves, Liang et al. (2009) found that the
shallow-to-normal decay behavior might be due to a reference time
effect.

It has long been speculated that long GRBs are associated with the
deaths of massive stars and hence supernovae (SNe) (Colgate 1974;
Woosley 1993; see Zhang \& M\'{e}sz\'{a}ros 2004; Piran 2005;
M\'{e}sz\'{a}ros 2006; Woosley \& Bloom 2006 for reviews). The
collapsar model is the most promising scenario, in which the GRB
jets are powered by the accretion of an accretion disk or a torus
fed by the fall-back material from the collapsar envelope (e.g.,
Popham et al. 1999; Narayan et al. 2001; Kohri \& Mineshige 2002; Di
Matteo et al. 2002; Kohri et al. 2005; Lee et al. 2005; Gu et al.
2006; Chen \& Beloborodov 2007; Liu et al. 2007; Kawanaka \&
Mineshige 2007; Janiuk et al. 2007; Janiuk \& Proga 2008). Kumar et
al. (2008a, b) proposed that the canonical lightcurves may be
produced by the mass-accretion of different layers of progenitor
stars. In the framework of their model, the X-ray emission of GRBs
may give insight into the properties of the progenitors. In this
paper we investigate the characteristics of the X-ray afterglow
lightcurves in the GRB rest frame and infer the properties of
progenitor stars with a sample of 33 GRBs based on the model of
Kumar et al. Our sample selection and the method are presented in \S
2. In \S 3, we give the correlations between the prompt gamma-rays
and the X-rays in the shallow decay segment. Inferred parameters of
progenitor stars are reported in \S 4. The results are summarized in
\S 5 with some discussion. Throughout, a concordance cosmology with
parameters $H_0 = 71$ km s$^{-1}$ Mpc$^{-1}$, $\Omega_M=0.30$, and
$\Omega_{\Lambda}=0.70$ are adopted.

\section{Data}
The XRT data are downloaded from the {\em Swift} data archive. The
HEAsoft packages, including Xspec, Xselect, Ximage, and {\em Swift}
data analysis tools, are used for the data reduction. We use an IDL
code developed by Zhang et al. (2007b) to automatically process the
XRT data for all the bursts detected by Swift/BAT with redshift
measurements up to October of 2008. Our sample includes only those
XRT lightcurves that have a clear initial steep decay segment, a
shallow decay segment, and a normal decay segment. We get a sample
of 33 GRBs. We fit the spectra accumulated in the steep and shallow
segments with an absorbed power law model and derived their spectral
indices\footnote{Although the steep decay segment has significant
spectral evolution (Zhang et al. 2007b), we derive only the
time-integrated spectral index for our analysis.}. Regarding the
steep decay segment as a GRB tail due to the curvature effect (e.g.,
Liang et al. 2006, Zhang et al. 2007a, 2009; Qin 2009), we estimate
the time of last emission episode of the GRB phase with the relation
$\alpha=2+\beta$ (Kumar et al. 2000; Liang et al. 2006). We fit the
steep-to-shallow decay segment with,
\begin{equation}
F=F_0[(\frac{t-t_1}{t_1})^{-(2+\beta_1)}+(\frac{t_2-t_1}{t_1})^{-(2+\beta_1)}\times
(\frac{t}{t_2})^{-\alpha_2}], \label{k}
\end{equation}
where $t_1$ is zero time point of the last emission epoch of the
prompt gamma-rays and $t_2$ is the starting time of the shallow
decay segment. The end time of the shallow decay segment ($t_3$) is
taken as the break time between the shallow to normal decay phases.
Flares in the steep-to-shallow decay segments are removed, if any.
Technically, the shallow decay segments are poorly sampled for some
GRBs. We fix the $\alpha_2$ value in order to get a reasonable fit.
Illustrations of our fitting results for 24 bursts of our sample are
shown in Figure. 1. We derive the X-ray fluence $S_X$ in the time
interval [$t_2$, $t_3$] in the XRT band and calculate the isotropic
X-ray energy with $E_{\rm iso, X}=4\pi D_L^2 S_X/(1+z)$, where $D_L$
is the luminosity distance. The isotropic peak fluxes of the prompt
gamma-rays ($L_{\rm peak, \gamma}$) are in 1024 ms timescale. We
take the X-ray luminosity at $t_2$, $L_{\rm X, t_{2}}$, as a
characteristic luminosity of the shallow decay segment. Our results
are summarized in Table 1. With the data reported in Table 1, we
show the distributions of $t_1$, $t_2$, and $t_3$ in comparison with
GRB duration $T_{90}$ in Figure 2. It is found that the distribution
of $t_1$ is comparable to $T_{90}$, $t_2$ is about $100-1000$
seconds, and $t_3$ is in $10^4-10^5$ seconds.

\section{Correlations}
The correlations  between the prompt gamma-rays and the X-rays in
the shallow decay segment may reveal some physical relations between
these two phases. We show the pair correlation of the observables
between the two phases in Figure 3, and measure these correlations
with the Spearman correlation analysis. Our results are reported in
Table 2. We find that there are several outliers at $T_{90}<30$s in
the correlation of $T_{90}$ and $t_1$. It seems natural that for
long bursts $t_1$ could be a mark of the end of the prompt emission
epoch and will be very likely approximately equal to $T_{90}$, since
the time of the last pulse should occur close to the end of the
overall emission. This breaks down for shorter bursts, where the
offset between $T_{90}$ and $t_1$ becomes important. We also find
that $T_{90}$ is not correlated with the time intervals $t_2-t_1$
and $t_3-t_2$, indicating that the break features in the XRT
lightcurves are independent of the durations of the prompt
gamma-rays. The energy releases in the two phases are strongly
correlated. The $L_{\rm peak, \gamma}$ and $E_{\rm iso, \gamma}$ are
correlated with the X-ray luminosity at $t_2$ ($L_{\rm X, t_{2}}$).
This fact likely suggests that the two phases may have related
energy budgets from the same central engine, and the physical
conditions to power the gamma-rays and the X-rays should be similar.
\section{Properties of Progenitor Stars}
Kumar et al. (2008a, b) proposed that the prompt gamma-rays and
X-ray afterglows are due to accretion of different layers of a
collapsar by a newly-formed black hole (BH) with 10$M_{\odot}$. In
their model, the highly variable lightcurves of the prompt
gamma-rays are explained as production of the accretion of the
dense, clumpy materials of the stellar core, and the power-law decay
X-rays may be due to the accretion of the fall-back materials of the
progenitor envelope. In this section, we derive the properties of
the progenitor stars based on the model of Kumar et al.

As mentioned by Kumar et al. (2008b) and from the numerical
simulations from GRMHD by McKinney (2005), the efficiency of the
accreted energy to the radiation is likely to depend on many
details. Here we assume a uniform radiation efficiency of $1\%$ of
the accreted mass by a rotating BH (McKinney 2005). We also do not
consider the beaming effect. Then the masses accreted by the BH
during the prompt gamma-ray phase ($M_{\gamma}$) and during the
shallow decay phase ($M_{\rm X}$) are estimated with $M_{\rm
acc}\sim 100E_{\rm iso}/c^2$. The average accretion rate thus can be
estimated with $\dot{M}\sim M_{\rm acc}/T_{\rm acc}$, where $T_{\rm
acc}$ is the accretion timescale in the rest frame. Considering the
fall-back of total accreted particles as free-fall, the radius $r$
for the fall-back time $T^{'}$ can be estimated with
\begin{equation}
r_{10}\sim 1.5{T_2^{'}}^{2/3}M^{1/3}_{BH,1},
\end{equation}
where  $r_{10}=r/10^{10}$ cm, $M_{BH,1}=M_{BH}/10 M_{\odot}$, and
$T^{'}_2=T^{'}/10^2$s. We assume $M_{BH,1}=1$ in this work. The
rotation rate $f_{\Omega}(r)$ of the fall-back material at radius
$r$ is defined as a ratio of the local angular velocity $\Omega(r)$
to the local Keplerian velocity $\Omega_k(r)$ of the material at
$r$,
\begin{equation}
f_{\Omega}(r)\equiv\frac{\Omega(r)}{\Omega_k(r)}.
\end{equation}
Considering the viscosity among accreted particles before they reach
the BH and combining with Eq. (2), one can obtain
\begin{equation}
f_{\Omega}(r)\propto
(\frac{t_{\mathrm{acc}}\alpha_{\mathrm{vis}}}{10T^{'}})^{1/3},
\end{equation}
where $\alpha_{\mathrm{vis}}$ is the viscous parameter and
$t_{\rm{acc}}$ is the viscous accretion timescale of the fall-back
material. Please note that the timescale $T_{\rm acc}$ is different
from $t_{\rm{acc}}$ as $t_{\rm{acc}}\sim 2/\alpha_{\rm{vis}}\Omega_k
$ is the viscous accretion time of the fall-back material after it
has circularized but $T_{\rm acc}$ is the accretion time for
fall-back material within different layers of the progenitor star,
which is given by the fall-back time without considering the
viscosity among the particles. Assuming that the observed flux is
proportional to the accretion rate and that the timescale
$t_{\rm{du}}$ of the decrease of mass fall-back rate from $f_2$ to
$f_1$ is much larger than $t_{\rm{acc}}$, the upper limit of
$f_{\Omega}(r)$ can be obtained by (Kumar et al. (2008a, b)
\begin{equation}
(\frac{t_{\rm{du}}}{t_{\rm{acc}}})^2 \geq \frac{f_2}{f_1}.
\end{equation}

A lower limit on $f_{\Omega}(r)$ is derived from the centrifugally
supporting condition that the fall-back material is able to form an
accretion disk at a radius $r_d \approx r[f_{\Omega}(r)]^2$, i.e.,
$r_d \geq 3R_g$, where $R_g\equiv GM_{\mathrm{BH}}/c^2$ ($c$ is the
velocity of the light). For a convective envelope, the density
profile at $r$ is $\rho \propto r^{-\delta}$, where $\delta$ is
determined by the slope of the shallow decay segment, with
$\delta=3(\alpha_2+1)/2$. The sharp decline in the steep decay
indicates that the density in the transition region decreases
sharply. In this region, $t_{\rm{acc}}\gg t_{\rm{du}}$, so the
accretion in this region can be ignored.

With the data of the bursts in our sample, we calculate the radii
for stellar core $r_c$, transient region $r_t$, and envelope region
$r_e$, the limits of spin parameter $f_{\Omega}$, the index of the
density profile $\delta$, the accreted masses ($M_{c}$ and $M_{e}$),
and the average accretion rates ($\dot{M}_{c}$ and $\dot{M}_{e}$) in
the prompt gamma-ray and the shallow decay phases. They are
tabulated in Table 3. We show the distributions of these parameters
in Figure 4. We find that the derived radii of the core layers  of
the progenitor stars for all the bursts are $r_c=10^9\sim 10^{10}$
cm with the rotation parameter as $f_{\Omega,c}=0.02\sim 0.05$. The
masses of the core layers for about two-thirds of GRBs in our sample
are $M_c=0.1\sim 5 M_\odot$ with a mass density of $10^{2}\sim
10^{5}$ g cm$^{-3}$, and their average accretion rates in the prompt
gamma-ray phase are $\dot{M}_c=0.01\sim 1 M_\odot$/s.

For the envelope layer, the estimated radii, lower limits on the
rotation parameters, and the masses of the fall-back materials are
$r_e=10^{10}\sim 10^{12}$ cm, the lower limit for the rotation
parameter $f_{\Omega,e}=10^{-3}\sim 10^{-2}$, and $M_e=10^{-3}\sim
1M_\odot$, respectively. The average accretion rates in the shallow
decay phase are much lower than that in the prompt gamma-ray phase,
i.e., $\dot{M}_e=10^{-8}\sim 10^{-4}M_\odot$/s, but they are
correlated. We measure the correlation with the Spearman correlation
analysis, which yields $\log \dot{M}_{c}=(-4.03\pm 0.21)+(0.78\pm
0.14)\log \dot{M}_{e}$ with a correlation coefficient $r=0.72$ and a
chance probability $p<10^{-4}$, as shown in Figure 5. The estimated
mass density in the envelope is $\sim 10^{-4}$ g cm$^{-3}$.

\section{Conclusions and Discussion}
We have investigated the characteristics of the X-rays in the GRB
rest frame, and inferred the properties of progenitor stars with a
sample of 33 GRBs based on the model of Kumar et al. Assuming that
the steep decay segment is due to the curvature effect, we fit the
lightcurves with a broken power-law to derive the zero time of the
last emission epoch of the prompt emission ($t_1$), the beginning
($t_2$) and the end time ($t_3$) of the shallow decay segment. The
$T_{90}$ is roughly consistent with $t_1$, but it is not correlated
with the time intervals of $t_2-t_1$ and $t_3-t_2$. The $E_{\rm iso,
\gamma}$ and $L_{\rm peak, \gamma}$ are correlated with $E_{\rm iso,
X}$ and $L_{X, t_{2}}$. This fact likely suggests that the energy
budgets  for the two phases may be from the same central engine.

Based on a model proposed by Kumar et al. (2008a, b), we inferred
the properties of the progenitor star with both the prompt
gamma-rays and the X-ray data. The derived radii of the core layers
of the progenitor stars for all the bursts are $r_c=10^8\sim
10^{10}$ cm with a rotation parameter as $f_{\Omega,c}=0.02\sim
0.05$. The masses of the core layers for about two-thirds of GRBs in
our sample are $M_c=0.1\sim 5 M_\odot$ with a mass density of
$10^{2}\sim 10^{5}$ g cm$^{-3}$, and their average accretion rates
in the prompt gamma-ray phase are $\dot{M}_c=0.01\sim 1 M_\odot$/s.
The estimated radii, lower limits on the rotation parameters, and
the masses of the fall-back materials for the envelope layers are
$r_e=10^{10}\sim 10^{12}$ cm, $f_{\Omega,e}=10^{-3}\sim 10^{-2}$,
and $M_e=10^{-3}\sim 1M_\odot$, respectively. The average accretion
rates in the shallow decay phase are much lower than those in the
prompt gamma-ray phase, i.e., $\dot{M}_e=10^{-8}\sim
10^{-4}M_\odot$/s, but they are correlated. The estimated mass
density in the envelope is $\sim 10^{-4}$ g cm$^{-3}$.

The connection between long-duration GRBs and SNe was predicted
theoretically (Colgate 1974; Woosley 1993) and has been verified
observationally through detecting spectroscopic features of the
underlying SNe in some nearby GRBs (Woosley \& Bloom 2006). The
collapsar model is the most promising scenario to explain the huge
release of energy associated with long duration GRBs (Woosley \&
Weaver 1995; Paczy\'{n}ski 1998; MacFadyen \& Woosley 1999; Zhang et
al. 2003; Janiuk \& Proga 2008). In this scenario GRBs are produced
by a jet powered by accretion of the core and the fall-back
materials of the progenitor star through a torus. We infer the
properties of the progenitor stars by assuming that both the prompt
gamma-rays and the X-rays observed with XRT are due to the accretion
of different layers of progenitor stars. We compare the distribution
of $r_e$, the radius of envelope region, with the photospheric radii
of a sample with 25 WC-type and 61 WN-type Wolf-Rayet (WR) stars
(Koesterke \& Hamann 1995, Li 2007) in Figure 6. We find that the
photospheric radius of a WR star is larger than $r_e$, consistent
with our prediction.

In our calculation, we regard that all GRBs are from a unified
collapsar with a central BH of $M=10M_\odot$. Our results thus might
be compared with simulation results for a collapsar with a given
mass. We compare the derived mass density profile as a function of
radius $r$ with simulations in Figure 7, in which the simulated mass
density profile is taken from Woosley \& Weaver (1995) for a massive
star with $M=25M_\odot$ (see also Janiuk \& Proga 2008). It is found
that, although the derived $\rho$ are systematically larger than the
results of simulations, they are very consistent with the
simulations.

In the collapsar models, the accretion duration should be as long as
the material fall-back timescale from the collapsar envelope
available to fuel the accretion disk or torus. Rotation of the
progenitor star should be high enough to form  the disk or torus.
One thus might expect a relation between the burst duration and the
rotation parameter $f_{\Omega}$. For a progenitor star with higher
rotation, the angular momentum loss should be longer, and the
accretion timescale might be longer, hence a longer GRB event. We
show the correlation between $t_1$ and $f_{\Omega}$ in Figure 8. A
tentative correlation is found, with a correlation coefficient of
$0.72$ and a chance probability $p<10^{-4}$. This correlation
indicates that the higher $f_{\Omega}$, the longer GRB could be
observed, consistent with the model's expectation.

It is believed that GRBs are highly collimated, with a beaming
factor $f_b\sim 1/500$ from the optical afterglow observations(Frail
et al. 2001). The measurement of the beaming angle has also proven
exceedingly difficult in the {\em Swift} era (Cenko et al. 2009). In
our analysis, we do not considering the beaming effect. The lack of
detection of jet-like breaks in the late XRT lightcurve might
suggest that the X-ray jet would be less collimated than expected
from the optical data (Burrows \& Racusin 2006; Liang et al. 2008).
In spite of this, the accretion rates and the accreted masses could
be up to 2 orders of magnitude lower than those derived in this
analysis if beaming is considered.

We appreciate valuable suggestions and comments from the anonymous
referee. We also thank Bing Zhang, Li-Xin Li, Zhuo Li, and Tong Liu
for helpful discussion.

\clearpage
\begin{table*}
\caption[]{XRT Observations and our fitting results}
  \label{Tab:publ-works}
\begin{minipage}{\textwidth}
\begin{center}\begin{tabular}{cccccccccccccccccccccc}
  \hline%\noalign{\smallskip}
%aaa\footnote{test} & bbb & ccc \\
GRB &$S_x$\footnote{The X-ray fluence integrated from $t_2$ to $t_3$
and its error in the XRT band (0.3-10 keV)} &$\Gamma_x$\footnote{The
time-averaged photon index of the steep decay phase}&
${t_1}$\footnote{The zero time of the emission epoch corresponding
to the steep decay segment}  & $t_2$\footnote{$t_2$ and $t_3$ are
the begin and the end times of the shallow decay segment}& $t_3$
&$\alpha_2$\footnote{The
slopes of the shallow decay segment}&$E_{\rm iso, X}$&$L_{\rm X, t_2}$\\
  &($10^{-7}$ erg $\mathrm{cm}^{-2}$)  &  &$(\mathrm{s})$  & $(\mathrm{s})$    & (ks)  &  & ($10^{50}$ erg)&($10^{47}$erg/s)&\\

  \hline\noalign{\smallskip}
050416A &   0.62    $\pm$   0.38    &   2.15    &    $\sim79$     &    $\sim87$   &   1.74    &   0.70    &   0.7 $\pm$   0.4 &   3.7 \\
050803  &   5.96    $\pm$   0.51    &   1.88    &   104  $\pm$   5    &   263  $\pm$   11   &   13.71   &   0.36    &   2.6 $\pm$   0.2 &   0.8 \\
050908  &   0.13    $\pm$   0.11    &   3.90    &   120  $\pm$   50   &   684  $\pm$   82   &   8.00    &   1.01    &   2.9 $\pm$   2.4 &   1.4 \\
051016B &   2.18    $\pm$   1.10    &   2.82    &   50   $\pm$   5    &   157  $\pm$   12   &   6.64   &   0.14    &   4.9  $\pm$   2.5 &   0.6 \\
051109A &   3.46    $\pm$   0.75    &   2.33    &   62   $\pm$   23   &   173  $\pm$   33   &   7.30    &   0.42    &   42.9 $\pm$   9.3 &   109.3 \\
060108  &   0.53    $\pm$   0.17    &   1.91    &   40   $\pm$   30   &   186  $\pm$   31   &   22.08   &   0.39    &   5.1  $\pm$   1.6 &   5.4 \\
060210  &   4.86    $\pm$   0.69    &   1.93    &   298  $\pm$   8    &   452  $\pm$   11   &   7.00    &   0.80    &   141.0 $\pm$   20.0    &   721.2 \\
060418  &   1.38    $\pm$   0.66    &   2.04    &   81  $\pm$   2    &   309  $\pm$   4    &   1.00    &   $\sim 0$&   7.6 $\pm$   3.6 &   51.6 \\
060502A &   5.09    $\pm$   1.19    &   2.43    &   12   $\pm$   6    &   190  $\pm$   13   &   72.57   &   0.59    &   28.6 $\pm$   6.7 &   16.9 \\
060510B &   0.28    $\pm$   0.27    &   1.42    &   310  $\pm$   2    &       $\sim3205$       &   170.00  &   $\sim 0$&   11.4 $\pm$   10.9 &   0.7 \\
060522  &   0.12    $\pm$   0.20    &   1.97    &   117  $\pm$   15   &   248  $\pm$   16   &   0.73    &   $\sim 0$&   5.2 $\pm$   8.6 &   114.7 \\
060526  &   0.46    $\pm$   0.26    &   1.80    &   266  $\pm$   1    &   1023 $\pm$   18   &   10.00   &   $\sim 0$&   9.6 $\pm$   5.4 &   9.2 \\
060605  &   0.82    $\pm$   0.52    &   1.60    &   59   $\pm$   74   &   455  $\pm$   42   &   7.00    &   $\sim 0$&   22.8 $\pm$   14.5 &   29.3 \\
060607A &   8.45    $\pm$   0.17    &   1.79    &   214  $\pm$   12   &   384  $\pm$   10   &   12.34   &   0.44    &   166.0 $\pm$   3.3 &   408.0 \\
060707  &   0.55    $\pm$   0.26    &   2.00    &   56   $\pm$   22   &   505  $\pm$   76   &   10.00   &   0.39    &   12.8 $\pm$   6.0 &   12.5 \\
060708  &   0.96    $\pm$   1.06    &   2.51    &   20   $\pm$   2    &   231  $\pm$   19   &   6.66    &   0.39    &   11.5 $\pm$   12.7 &   16.8 \\
060714  &   1.48    $\pm$   0.46    &   2.02    &   145  $\pm$   4    &   311  $\pm$   15   &   3.70    &   0.02    &   23.5 $\pm$   7.3  &   33.1 \\
060729  &   19.58   $\pm$   0.83    &   2.71    &   120  $\pm$   2    &   425  $\pm$   8    &   72.97   &   0.27    &   14.3 $\pm$   0.6 &   1.0 \\
060814  &   6.93    $\pm$   0.87    &   1.84    &   81   $\pm$   13   &   967  $\pm$   74   &   17.45   &   0.15    &   12.5 $\pm$   1.6 &   1.4 \\
060906  &   0.96    $\pm$   0.29    &   2.44    &   85   $\pm$   26   &   222  $\pm$   21   &   13.66   &   0.33    &   25.2 $\pm$   7.6 &   20.7 \\
061121  &   19.89   $\pm$   6.14    &   1.62    &   103  $\pm$   2    &   176  $\pm$   3    &   2.43   &   0.25    &   85.4 $\pm$   26.4 &   60.5 \\
070110  &   3.59    $\pm$   0.23    &   2.11    &   61   $\pm$   3    &   522  $\pm$   19   &   20.40   &   0.17    &   44.7 $\pm$   2.9 &   17.2 \\
070306  &   2.53    $\pm$   0.94    &   2.29    &   110  $\pm$   2    &   542   $\pm$   8    &   15.00   &   $\sim 0$&   14.0  $\pm$   5.2 &   4.6 \\
070318  &   0.79    $\pm$   1.45    &   1.40    &   86   $\pm$   12   &   809  $\pm$   47   &   2.00    &   $\sim 0$&   1.4  $\pm$   2.6 &   3.2 \\
070721B &   1.80    $\pm$   1.38    &   1.48    &   289  $\pm$   9    &   450  $\pm$   28   &   7.50    &   0.62    &   46.2 $\pm$   35.4 &   192.2 \\
071021  &   0.24    $\pm$   0.39    &   2.12    &   175  $\pm$   3    &   558  $\pm$   74   &   20.00   &   0.37    &   10.1 $\pm$   16.6 &   15.6 \\
080310  &   1.19    $\pm$   0.53    &   1.45    &   504  $\pm$   1    &   $\sim1313$           &   20.00   &   0.33    &   15.7 $\pm$   7.0 &   9.1 \\
080430  &   0.82    $\pm$   0.23    &   2.42    &   0    $\pm$   5    &   165  $\pm$   12   &   8.80    &   0.52    &   1.2 $\pm$   0.3 &   2.2 \\
080607  &   2.85    $\pm$   0.81    &   1.81    &   100  $\pm$   2    &   238  $\pm$   7    &   1.50   &   1.03    &   54.7 $\pm$   15.5 &   954.8 \\
080707  &   0.24    $\pm$   0.13    &   2.10    &   34   $\pm$   8    &   192  $\pm$   28   &   7.60   &   0.16    &   0.9 $\pm$   0.5 &   8.4  \\
080905B &   3.50    $\pm$   2.34    &   1.49    &   62   $\pm$   5    &   179   $\pm$   7    &   6.50   &   $\sim 0$&   44.4 $\pm$   29.7 &   104.4 \\
081007  &   0.96    $\pm$   0.31    &   3.00    &             $\sim35$   &   188  $\pm$   5    &   40.00  &   0.69    &   0.7 $\pm$   0.2 &   1.4 \\
081008  &   1.59    $\pm$   0.52    &   1.91    &   232   $\pm$   3   &   484  $\pm$   15   &   20.00  &   0.73    &   14.5 $\pm$   4.7 &   35.0  \\
\hline
% \noalign{\smallskip}
\end{tabular}\end{center}
\end{minipage}
\end{table*}
%\clearpage
\begin{table*}
\caption[]{Spearman correlation coefficients between the prompt
gamma-rays and the X-rays in the shallow decay phase}
  \label{Tab:publ-works}
\begin{minipage}{\textwidth}
\begin{center}\begin{tabular}{cccccccccccccccccccccc}
\hline
  %\noalign{\smallskip}
 &$t_1$&$t_2-t_1$ &$t_3-t_2$&
$L_{\mathrm{X,t_2}}$ & $E_{\mathrm{iso,X}}$ \\
  \hline\noalign{\smallskip}
$T_{90}$ & 0.55(8.1E-4)\footnote{In the bracket is the chance probability.} & 0.58(2.9E-4) & 0.08(0.63) &0.27(0.13) &0.46(6.3E-3)   \\
$L_{\rm peak, \gamma}$ &   0.40(0.02) & 0.08(0.66)&  -0.30(0.08)    &  0.74($<$1E-4)   &    0.66($<$1E-4)   \\
$E_{\mathrm{iso}, \gamma}$  &   0.51(2.2E-3) & 0.33(0.06) & -0.18(0.31)  & 0.68($<$1E-4)  &   0.72($<$1E-4)     \\
\hline
% \noalign{\smallskip}
\end{tabular}\end{center}
\end{minipage}
\end{table*}

\clearpage

\begin{table*}
\caption[]{Inferred properties of the progenitor stars for the
bursts in our sample}
  \label{Tab:publ-works}
\begin{minipage}{\textwidth}
\begin{center}\begin{tabular}{cccccccccccccccccccccc}
  \hline%\noalign{\smallskip}
%aaa\footnote{test} & bbb & ccc \\
GRB & $r_c$\footnote{Radii of the core, transit and envelope region of the progenitor stars.} &$r_t^a$ & $r_e^a$&
$f_{\Omega,c}$\footnote{The range of the rotational parameter of the core layer,
$f_{\Omega}\equiv\frac{\Omega}{\Omega_k}$.} & $f_{\Omega,e,low}$\footnote{The lower limit on the rotational parameter
of the envelope layer.}&$\delta$\footnote{The density profile in the stellar
envelope: $\rho(r)\sim r^{-\delta}$.}  & $M_{c}$\footnote{The fall-back mass during the prompt phase.} & $M_{e}$\footnote{The fall-back mass during the shallow decay phase.}\\
      &    $(10^9 \mathrm{cm})$ &$(10^{10} \mathrm{cm})$ & $(10^{11} \mathrm{cm})$& & $(10^{-3})$ & & $M_{\odot}$ & 0.1$M_{\odot}$& \\
\hline\noalign{\smallskip}
050416A &   9.14 &   0.98 &   0.72 &   [0.21 0.02] &   7.85 &   2.56 &   0.02  &   0.04 $\pm$   0.02 \\
050803  &   2.15$\pm$1.57 &   2.26$\pm$0.27 &   3.16 &   [0.34 0.02] &   3.75 &   2.04 &   0.09 $\pm$  0.01 &   0.14 $\pm$   0.01 \\
050908  &   6.34$\pm$3.53 &   2.03$\pm$0.49 &   1.05 &   [0.19 0.03] &   6.52 &   3.02 &   0.61 $\pm$   0.06&   0.16 $\pm$   0.13 \\
051016B &   6.05$\pm$1.30 &   1.30$\pm$0.23 &   7.34 &   [0.29 0.03] &   2.46 &   1.71 &   0.02 $\pm$  0.01 &   0.27 $\pm$   0.14 \\
051109A &   4.86$\pm$2.50 &   0.97$\pm$0.32 &   1.17 &   [0.33 0.03] &   6.16 &   2.13 &   0.18$\pm$   0.03 &   2.38 $\pm$   0.51 \\
060108  &   3.87$\pm$3.23 &   1.08$\pm$0.33 &   2.62 &   [0.34 0.03] &   4.12 &   2.09 &   0.20 $\pm$   0.02&   0.28 $\pm$   0.09 \\
060210  &   10.74$\pm$0.93&   1.42$\pm$0.12 &   0.88 &   [0.30 0.02] &  7.10  &   2.70 &   12.29$\pm$   0.66&   7.81 $\pm$   1.11 \\
060418  &   7.09$\pm$0.50 &   1.73$\pm$0.10 &   0.38 &   [0.33 0.03] &  10.83 &   1.50 &   2.53 $\pm$  0.08 &   0.42 $\pm$   0.20 \\
060502A &   2.01$\pm$1.29 &   1.25$\pm$0.21 &   6.56 &   [0.24 0.05] &   2.60 &   2.39 &   0.69$\pm$   0.03 &   1.59 $\pm$   0.37 \\
060510B &   9.76$\pm$0.32 &   4.64 &   6.54 &   [0.38 0.02] &   2.61 &   1.50 &   9.27 $\pm$   0.40&   0.63 $\pm$   0.60 \\
060522  &   4.98$\pm$1.28 &   0.82$\pm$0.13 &   0.17 &   [0.33 0.03] &   16.26&   1.50 &   2.77$\pm$   0.27 &   0.29 $\pm$   0.47  \\
060526  &   11.04$\pm$0.30&   2.71$\pm$0.18 &   1.24 &   [0.35 0.02] &   5.99 &   1.50 &   1.47$\pm$   0.19 &   0.53 $\pm$   0.30 \\
060605  &   3.70$\pm$4.30 &   1.45$\pm$0.30 &   0.90 &   [0.36 0.03] &   7.05 &   1.50 &   1.07 $\pm$   0.14&   1.26 $\pm$   0.80 \\
060607A &   9.76$\pm$1.46 &   1.44$\pm$0.13 &   1.46 &   [0.34 0.02] &   5.53 &   2.15 &   2.84 $\pm$   0.11&   9.23 $\pm$   0.19 \\
060707  &   3.76$\pm$2.02 &   1.64$\pm$0.46 &   1.20 &   [0.31 0.03] &   6.09 &   2.09 &   2.08 $\pm$   0.20&   0.71 $\pm$   0.33 \\
060708  &   2.33$\pm$0.57 &   1.18$\pm$0.23 &   1.11 &   [0.24 0.04] &   6.32 &   2.09 &   0.33$\pm$   0.03 &   0.64 $\pm$   0.71 \\
060714  &   8.01$\pm$0.76 &   1.33$\pm$0.18 &   0.69 &   [0.33 0.02] &   8.00 &   1.53 &   2.65 $\pm$ 0.18  &   1.31 $\pm$   0.41 \\
060729  &   12.71$\pm$0.91&   2.95$\pm$0.20 &   9.12 &   [0.29 0.02] &   2.21 &   1.91 &   0.11   $\pm$ 0.01&   0.79 $\pm$   0.03 \\
060814  &   8.68$\pm$2.55 &   4.53$\pm$0.82 &   3.12 &   [0.32 0.02] &   3.78 &   1.72 &   1.50 $\pm$   0.02&   0.69  $\pm$   0.09 \\
060906  &   4.80$\pm$2.21 &   0.91$\pm$0.19 &   1.42 &   [0.31 0.03] &   5.59 &   1.99 &   3.22   $\pm$0.20 &   1.40 $\pm$   0.42 \\
061121  &   8.77$\pm$0.61 &   1.25$\pm$0.09 &   3.34 &   [0.33 0.02] &   3.65 &   1.88 &   3.27 $\pm$   0.05&   4.75 $\pm$   1.47  \\
070110  &   4.82$\pm$0.58 &   2.02$\pm$0.22 &   2.32 &   [0.30 0.03] &   4.38 &   1.75 &   1.11 $\pm$   0.07&   2.48 $\pm$   0.16  \\
070306  &   8.67$\pm$0.54 &   2.51$\pm$0.15 &   2.30 &   [0.31 0.02] &   4.40 &   1.50 &   1.65 $\pm$   0.09&   0.78 $\pm$   0.29  \\
070318  &   9.03$\pm$2.41 &   4.03$\pm$0.61 &   0.74 &   [0.39 0.02] &   7.77 &   1.50 &   0.25 $\pm$   0.01&   0.08 $\pm$   0.01 \\
070721B &   10.95$\pm$1.08&   1.47$\pm$0.23 &   0.96 &   [0.32 0.02] &   6.80 &   2.43 &   5.13 $\pm$   0.28&   2.57 $\pm$   1.96 \\
071021  &   6.61$\pm$0.42 &   1.43$\pm$0.37 &   1.55 &   [0.33 0.03] &   5.35 &   2.06 &   3.05 $\pm$   0.47&   0.56 $\pm$   0.09 \\
080310  &   19.39$\pm$0.39&   3.67 &   2.26 &   [0.37 0.02] &   4.44 &   2.00 &   1.68$\pm$   0.15 &   0.87 $\pm$   0.39 \\
080430  &   0.25$\pm$1.45 &   1.43$\pm$0.25 &   2.03 &   [0.11 0.03] &   4.68  &  2.28 &   0.10 $\pm$  0.01 &   0.07 $\pm$   0.02 \\
080607  &   5.92$\pm$0.38 &   1.05$\pm$0.10 &   0.36 &   [0.34 0.03]  &  11.12 &   3.05&   25.60 &   3.04 $\pm$   0.86 \\
080707  &   4.27$\pm$1.60 &   1.36$\pm$0.37 &   1.58 &   [0.32 0.03] &   5.31 &   1.74 &   0.11 $\pm$   0.01&   0.05 $\pm$   0.03 \\
080905B &   4.84$\pm$0.96 &   0.98$\pm$0.11 &   0.78 &   [0.37 0.03] &   7.55 &   1.50 &   1.27 $\pm$   0.14&   2.47 $\pm$   1.65 \\
081007  &   5.61 &   1.72$\pm$0.15 &   2.43 &   [0.25 0.03] &   4.27 &   2.54 &   0.03 $\pm$  0.01 &   0.04 $\pm$   0.01  \\
081008  &   12.74$\pm$0.70&   2.08$\pm$0.21 &   1.56 &   [0.33 0.02] &   5.33 &   2.60 &   2.18 $\pm$  0.10 &   0.81 $\pm$   0.26 \\

\hline
% \noalign{\smallskip}
\end{tabular}\end{center}
\end{minipage}
\end{table*}

\clearpage
\begin{figure*}
\includegraphics[angle=0,scale=0.35]{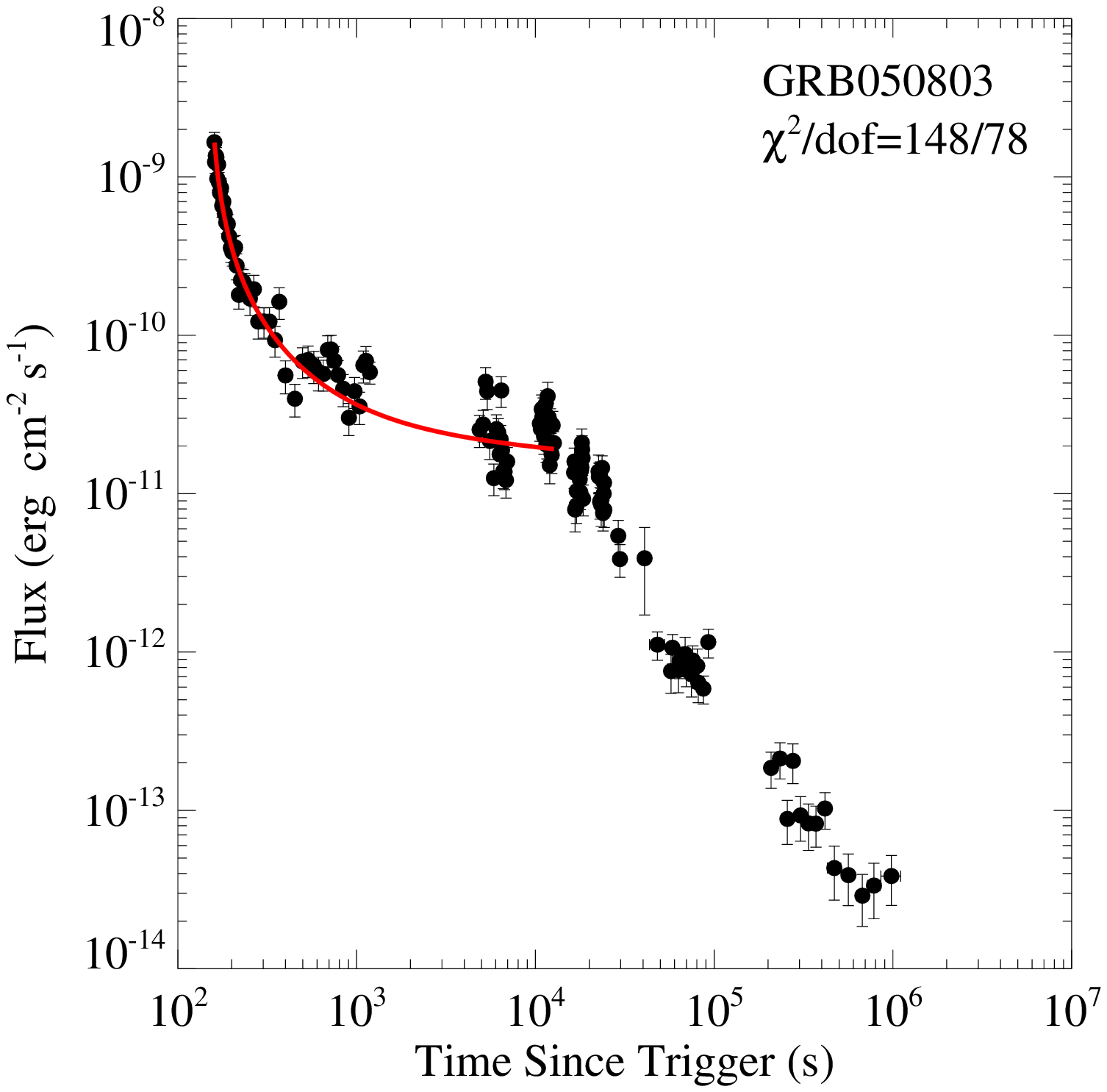}
\includegraphics[angle=0,scale=0.35]{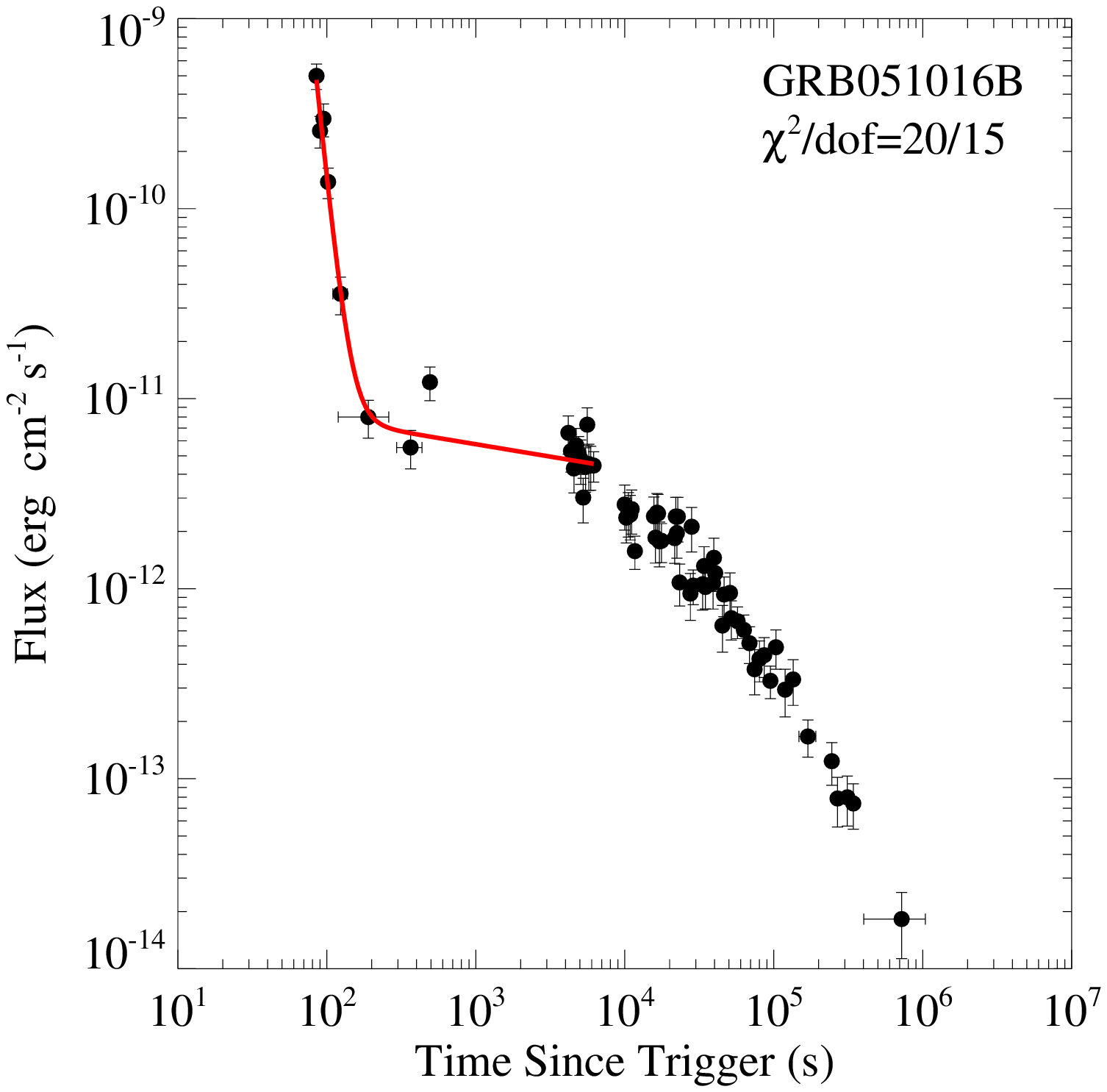}
\includegraphics[angle=0,scale=0.35]{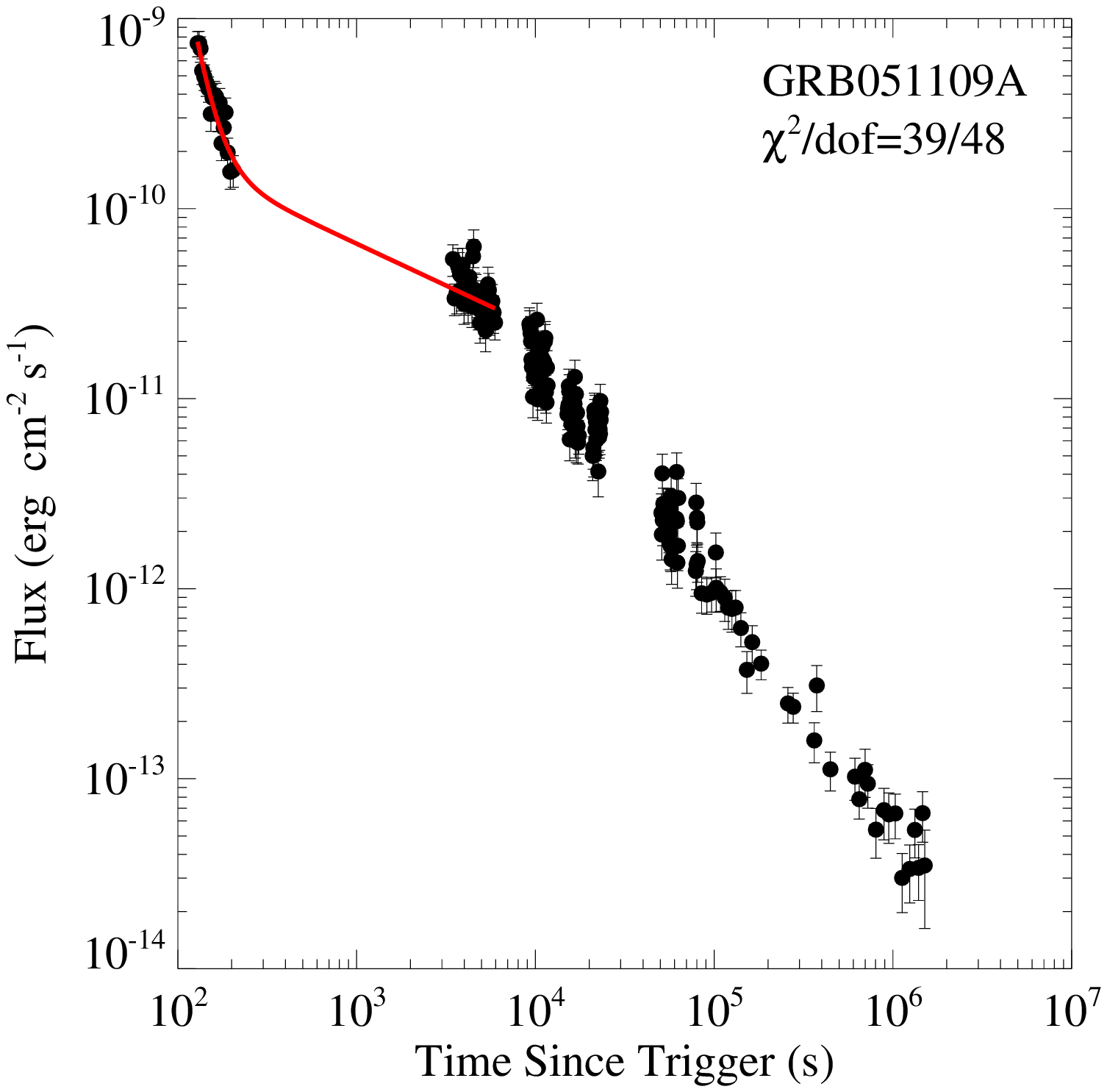}
\includegraphics[angle=0,scale=0.35]{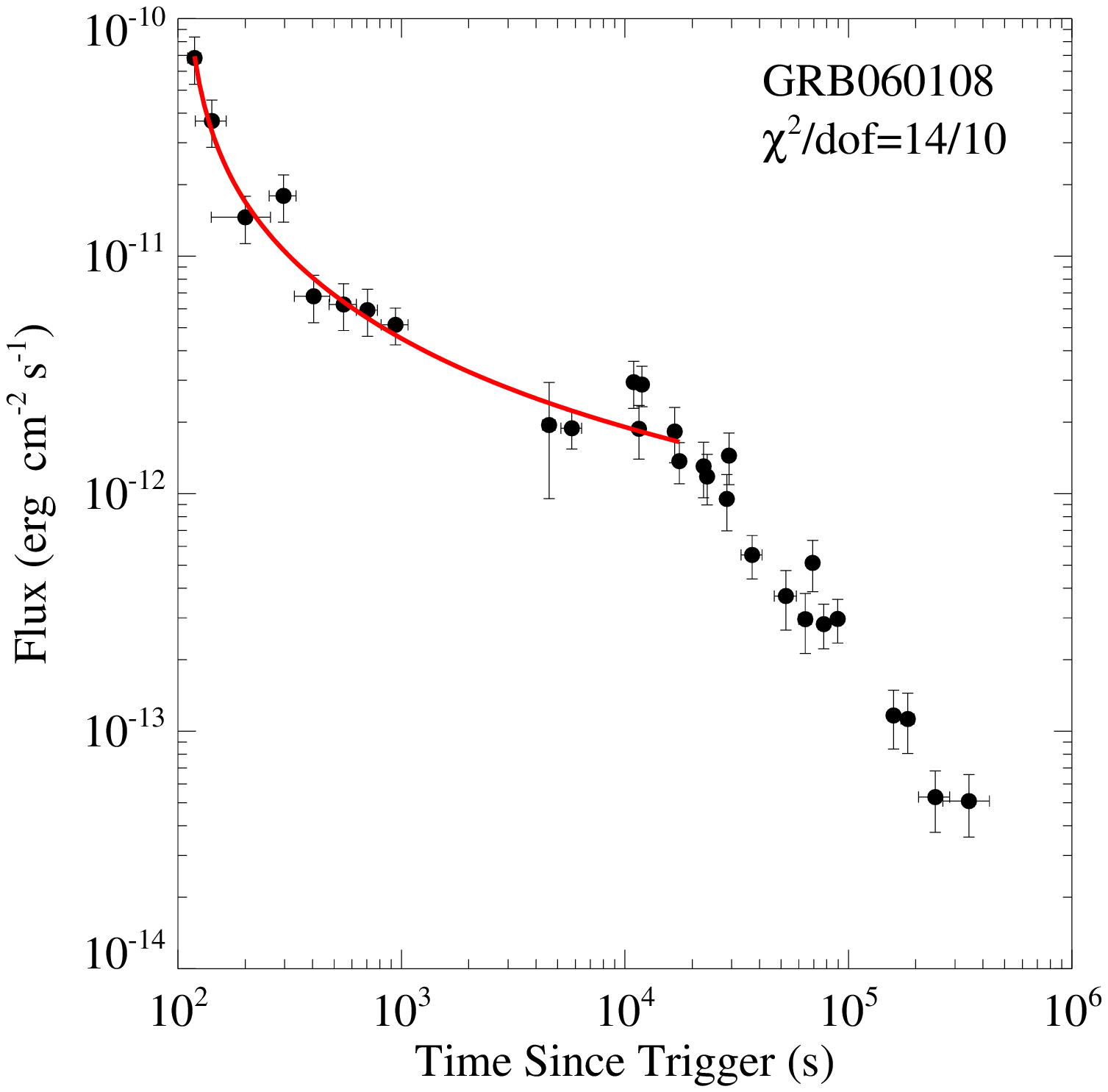}
\includegraphics[angle=0,scale=0.35]{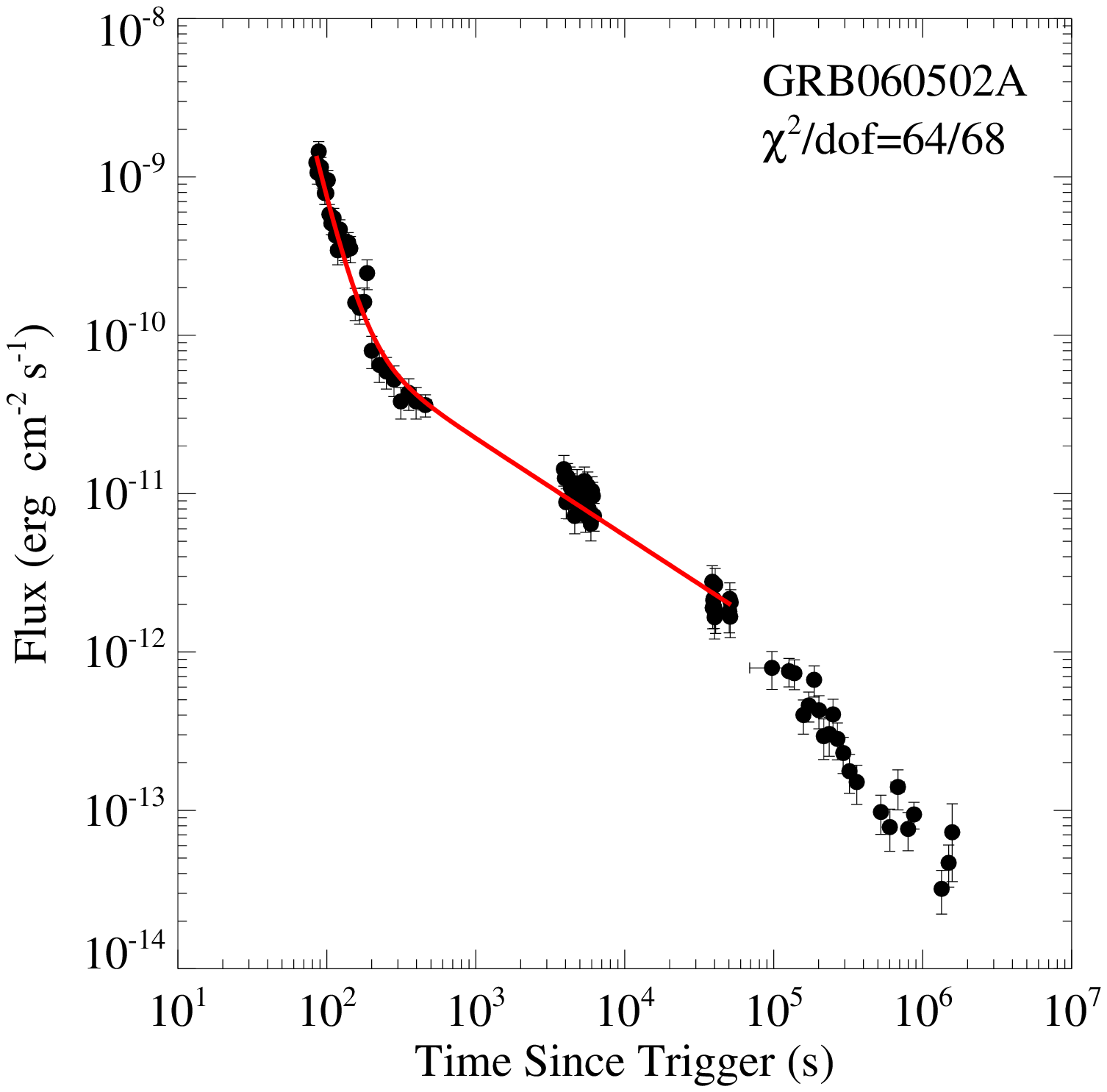}
\includegraphics[angle=0,scale=0.35]{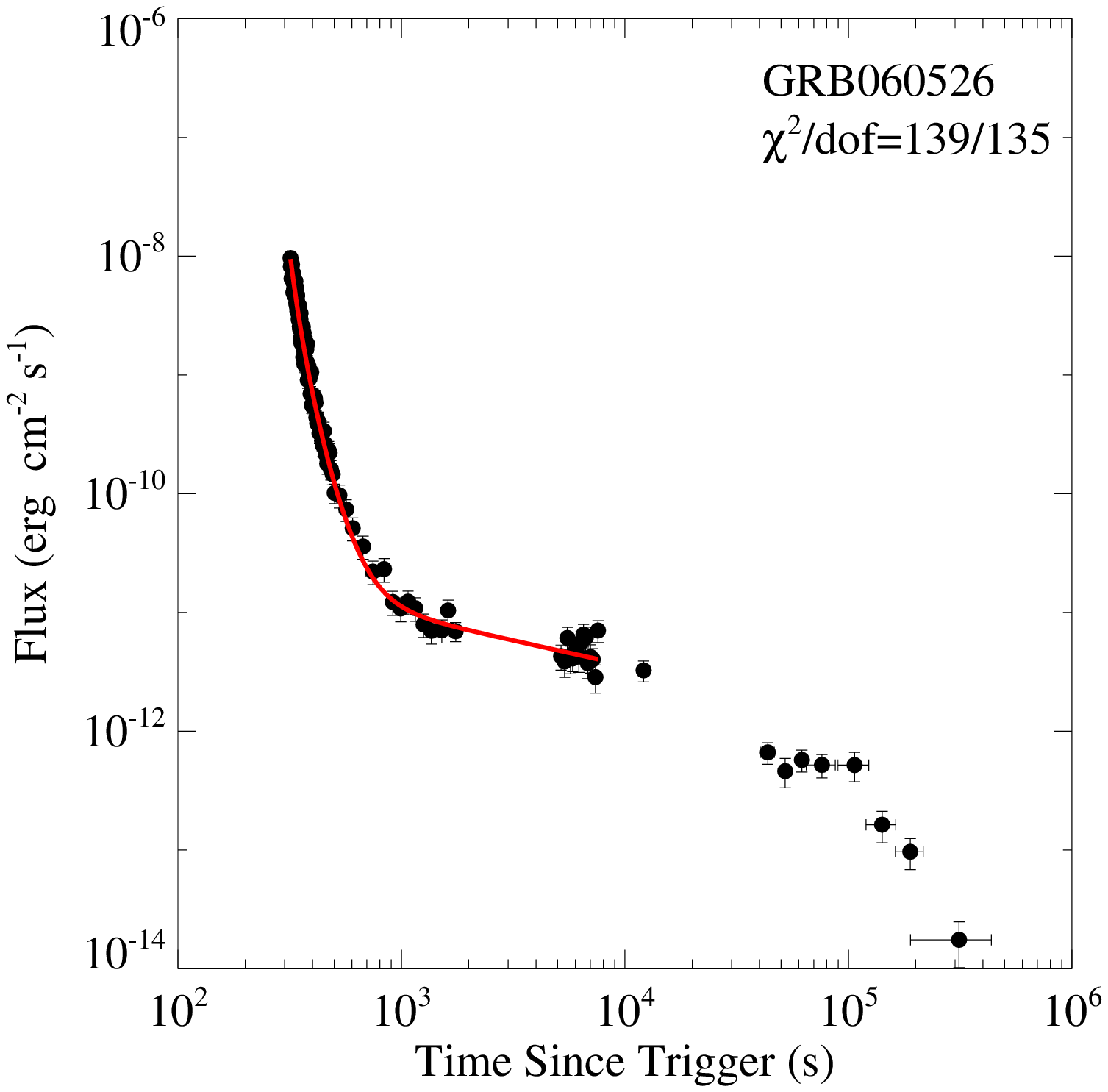}
\includegraphics[angle=0,scale=0.35]{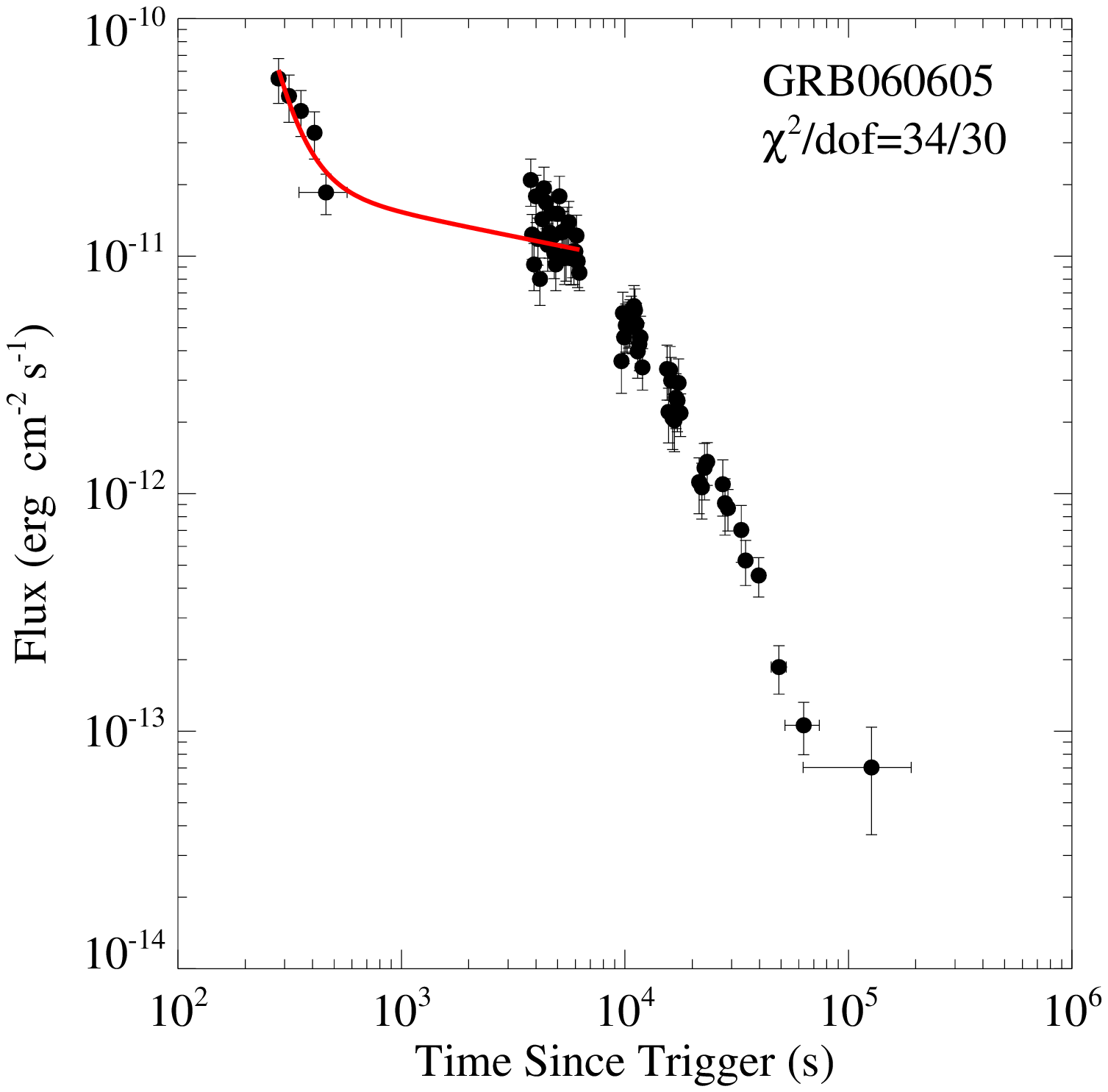}
\includegraphics[angle=0,scale=0.35]{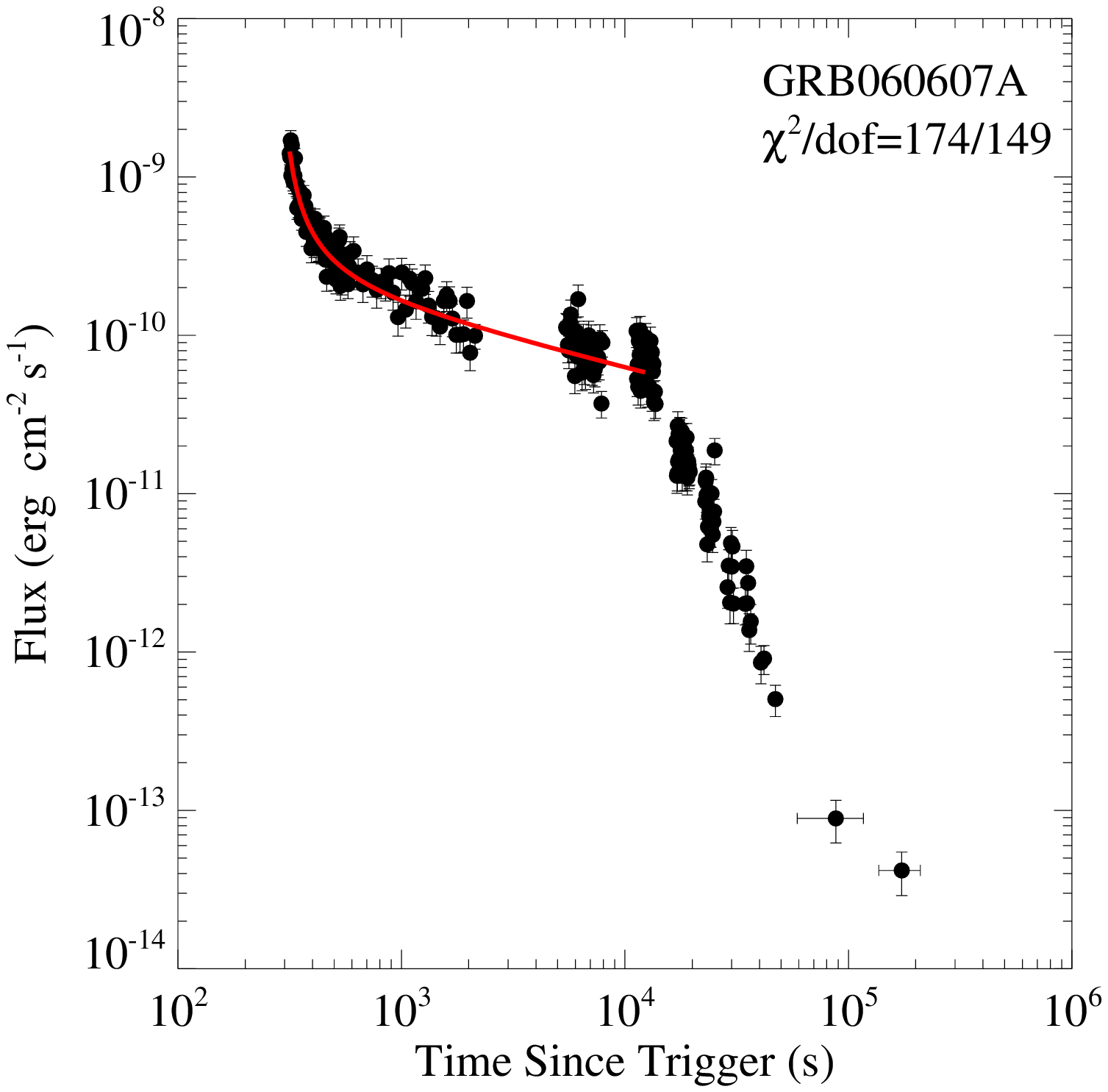}
\includegraphics[angle=0,scale=0.35]{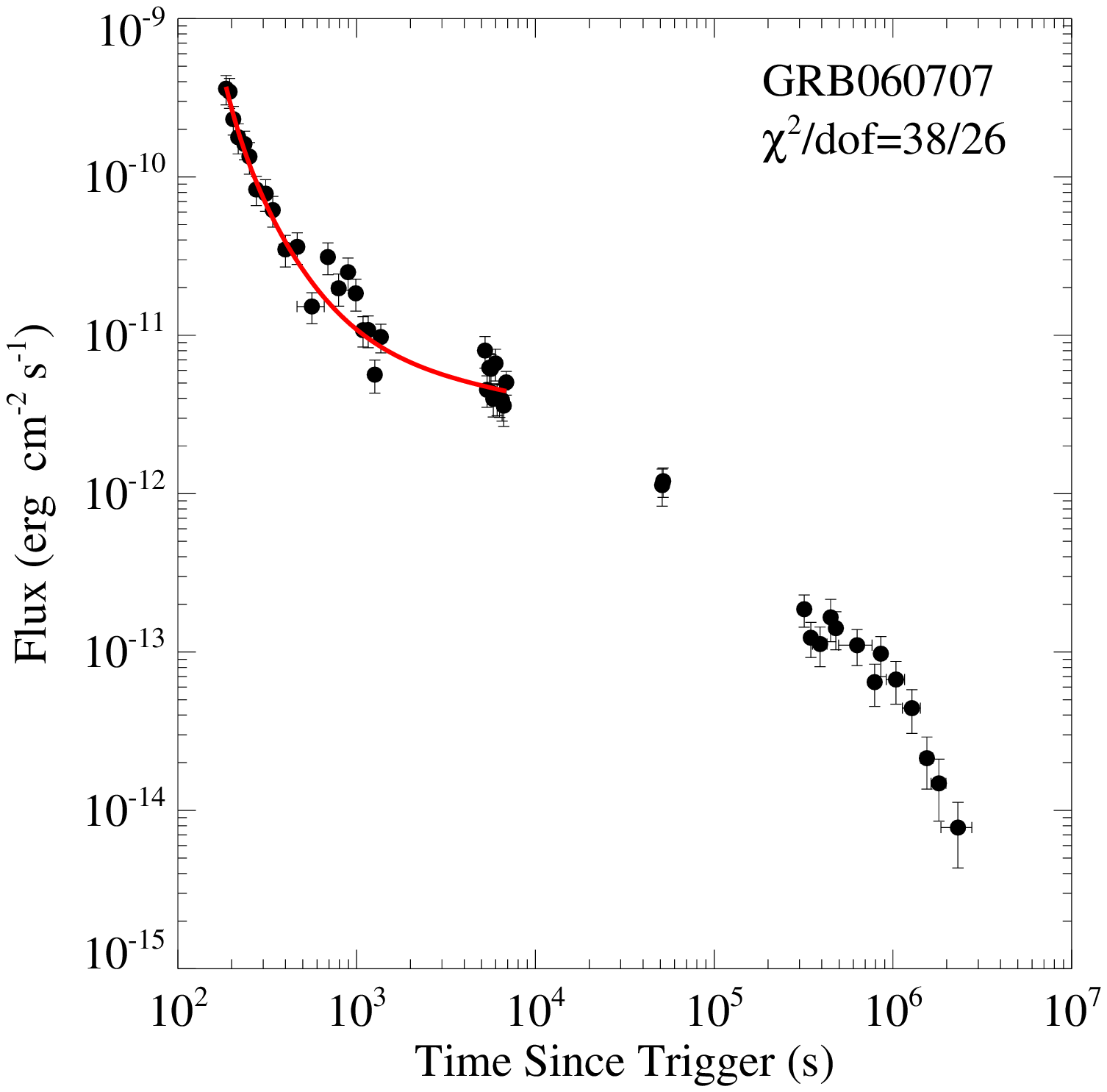}
\includegraphics[angle=0,scale=0.35]{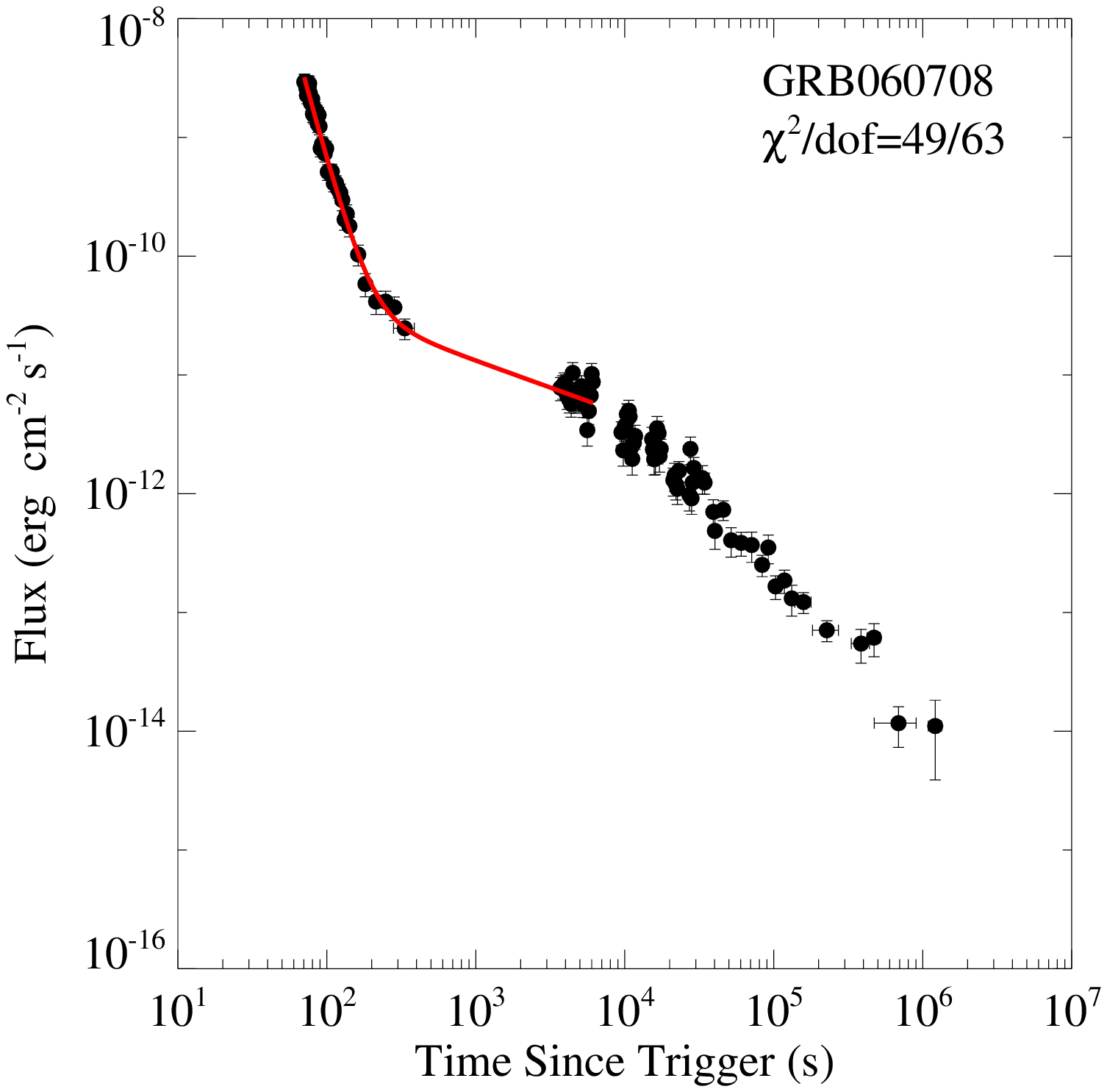}
\includegraphics[angle=0,scale=0.35]{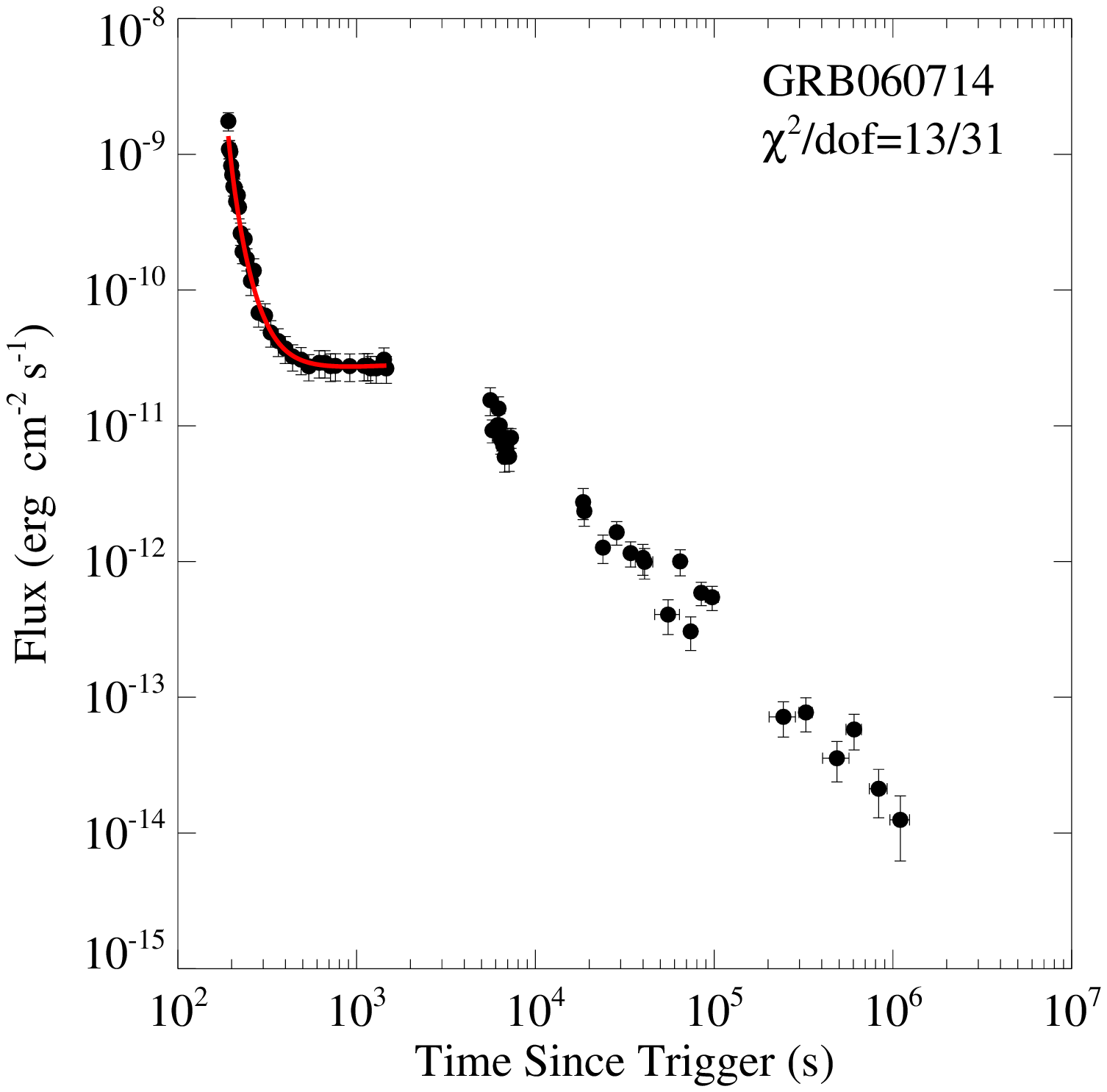}
\includegraphics[angle=0,scale=0.35]{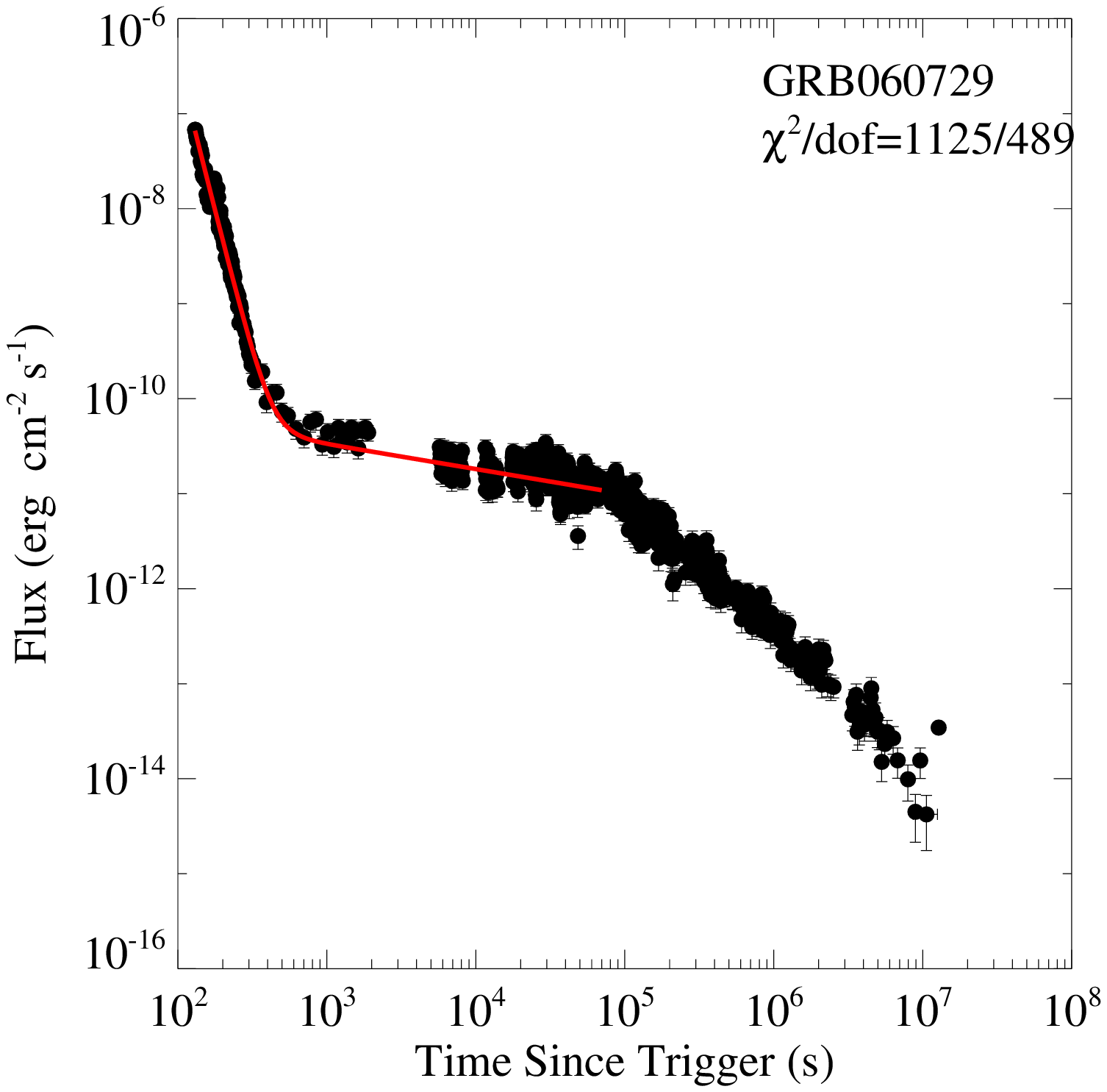}

\hfill \center{Fig.1  (continued)} \caption{Illustration of the XRT
light curves ({\em dots} with error bars) with our best fit {\em red
solid} for some bursts in our sample. The $\chi^2$ and degrees of
freedom of the fits are also marked in each panel.}\label{XRT_LC}
\end{figure*}

\clearpage
\begin{figure*}

\includegraphics[angle=0,scale=0.35]{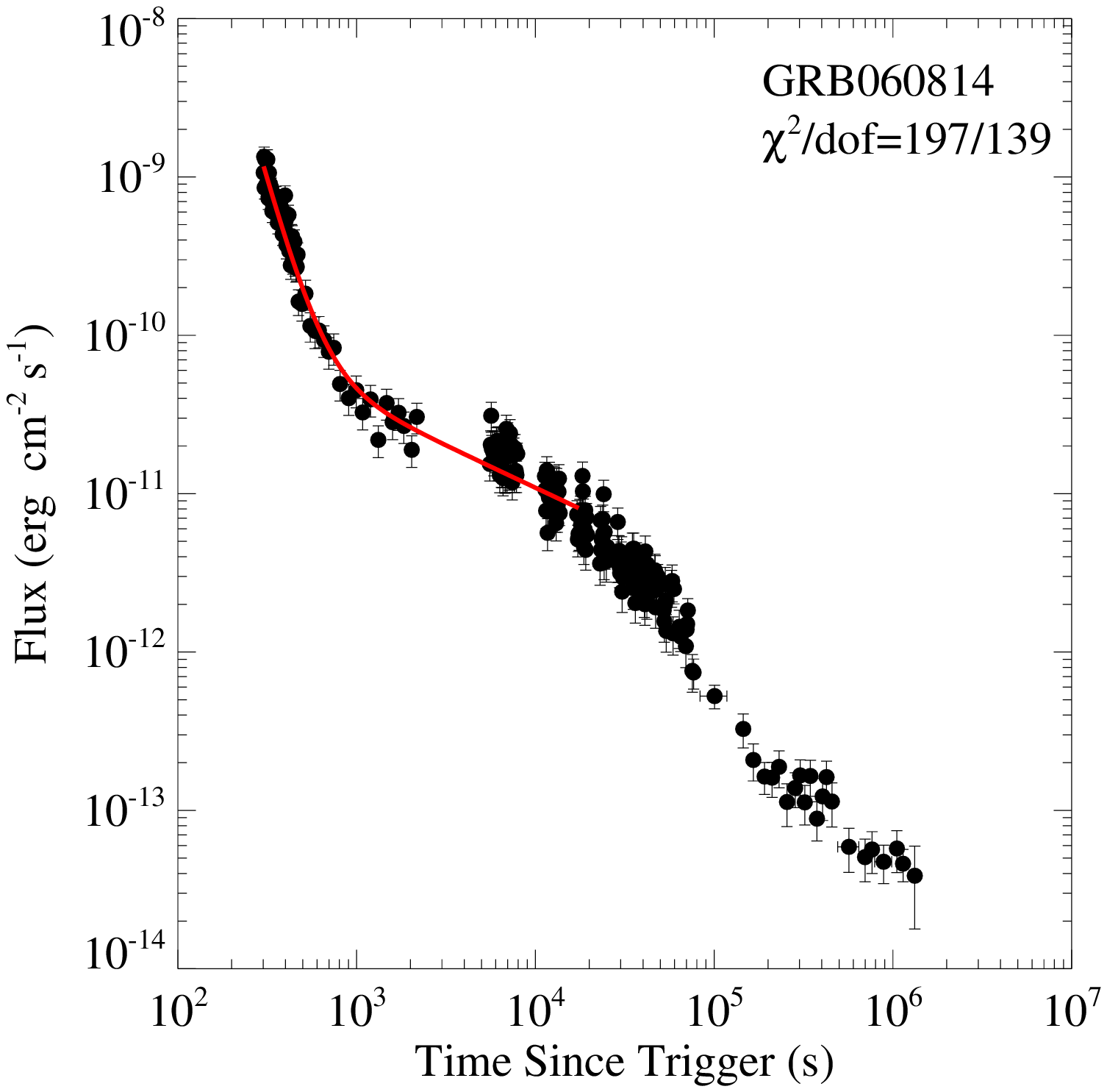}
\includegraphics[angle=0,scale=0.35]{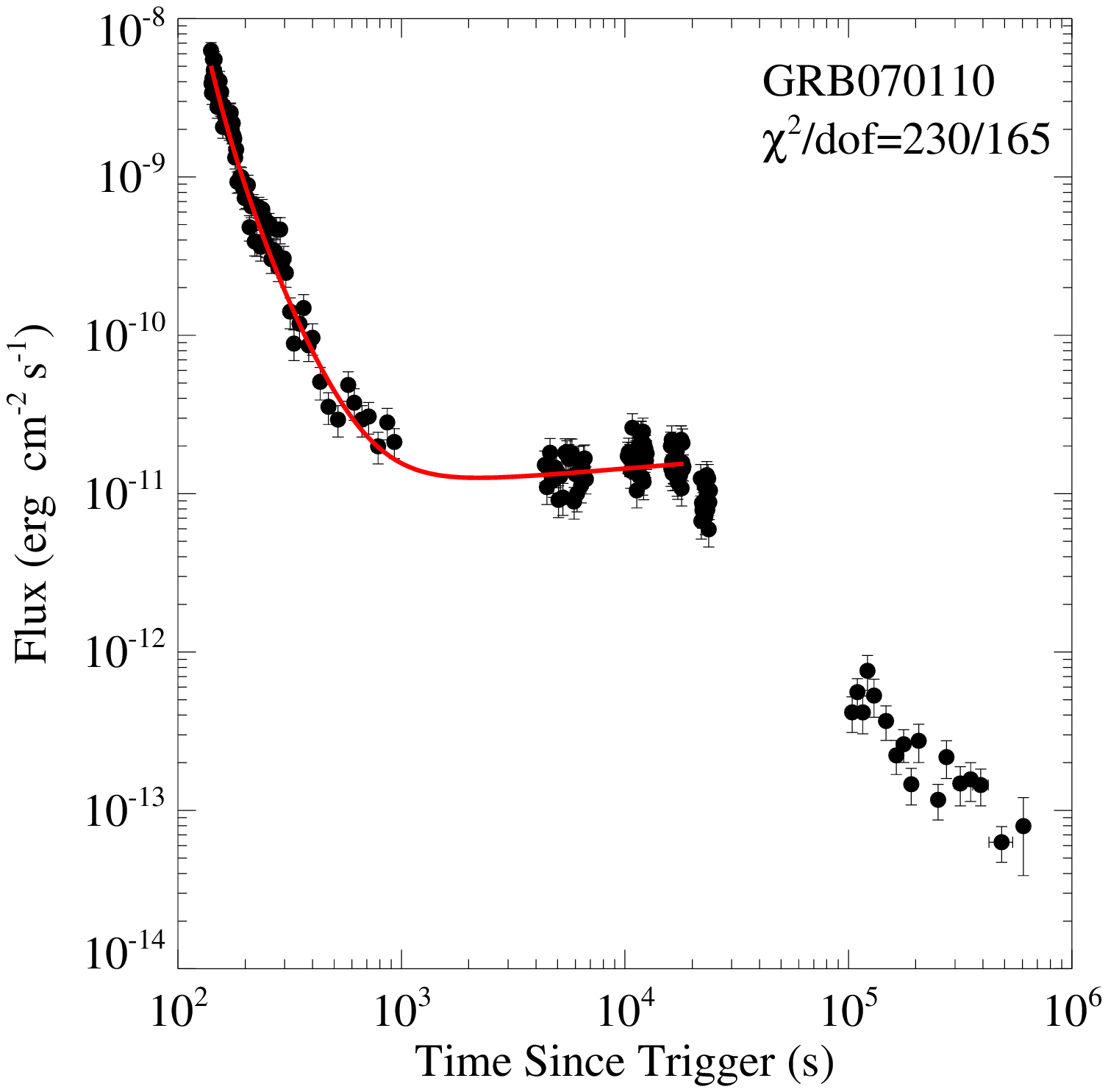}
\includegraphics[angle=0,scale=0.35]{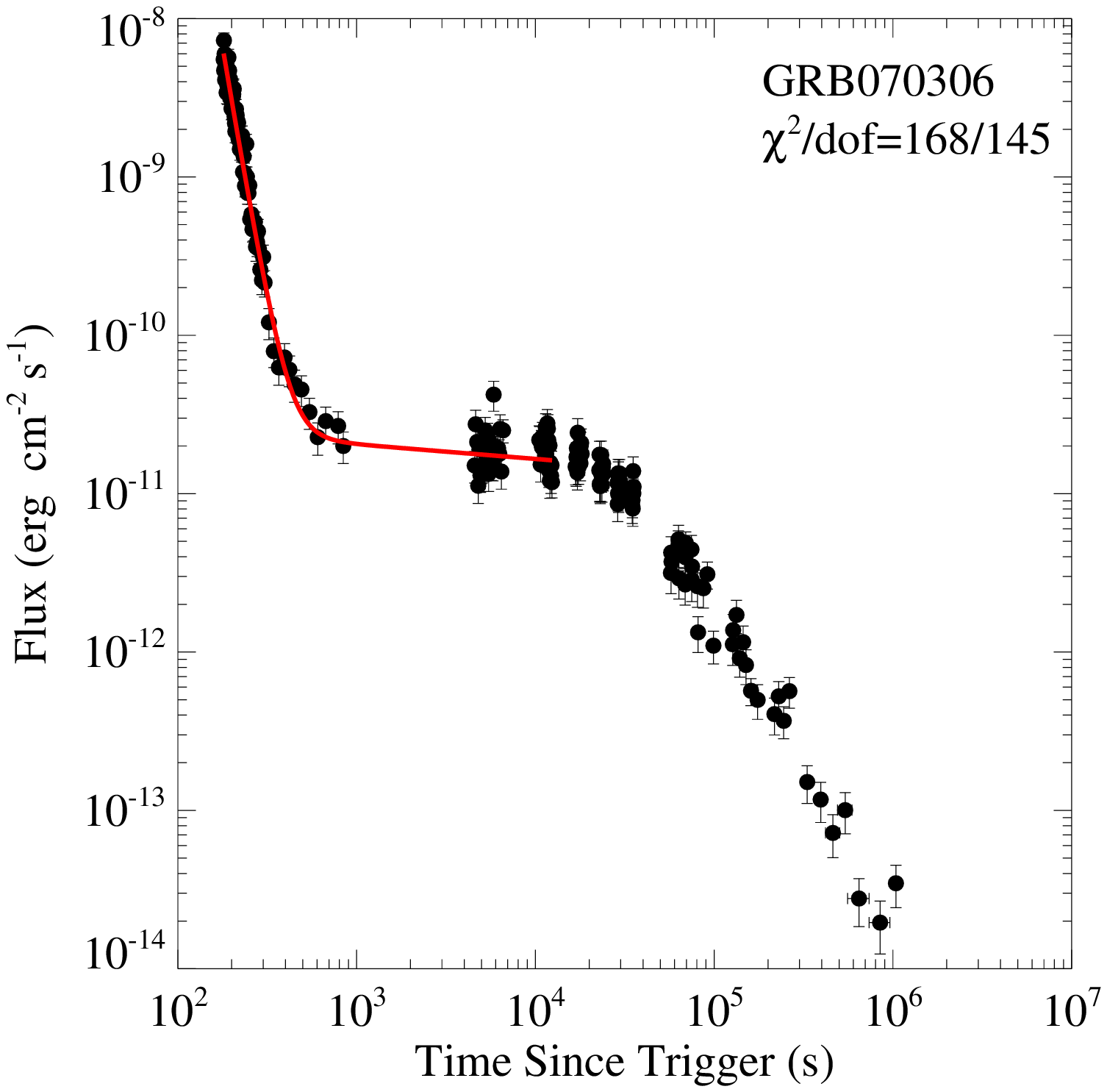}
\includegraphics[angle=0,scale=0.35]{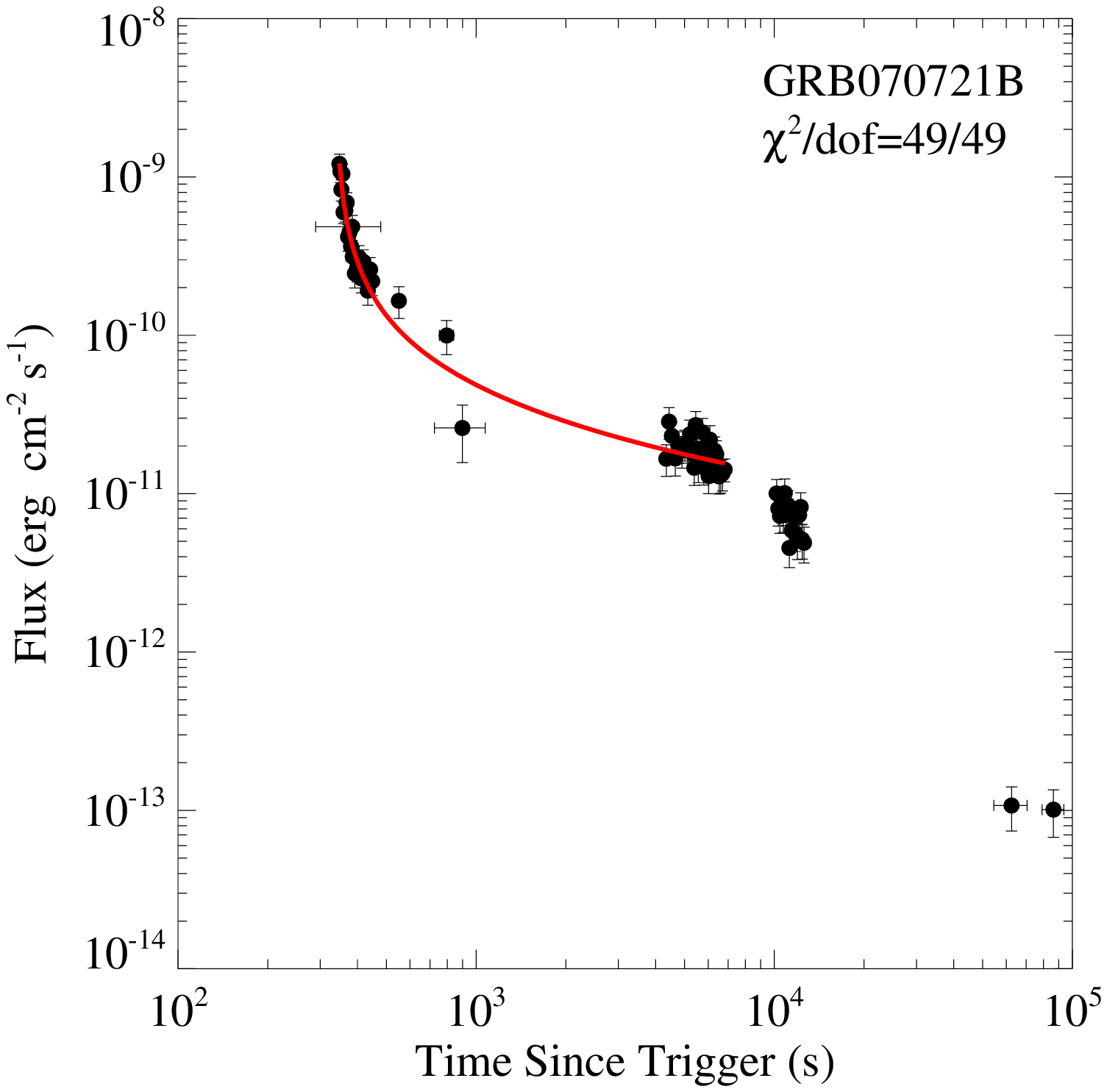}
\includegraphics[angle=0,scale=0.35]{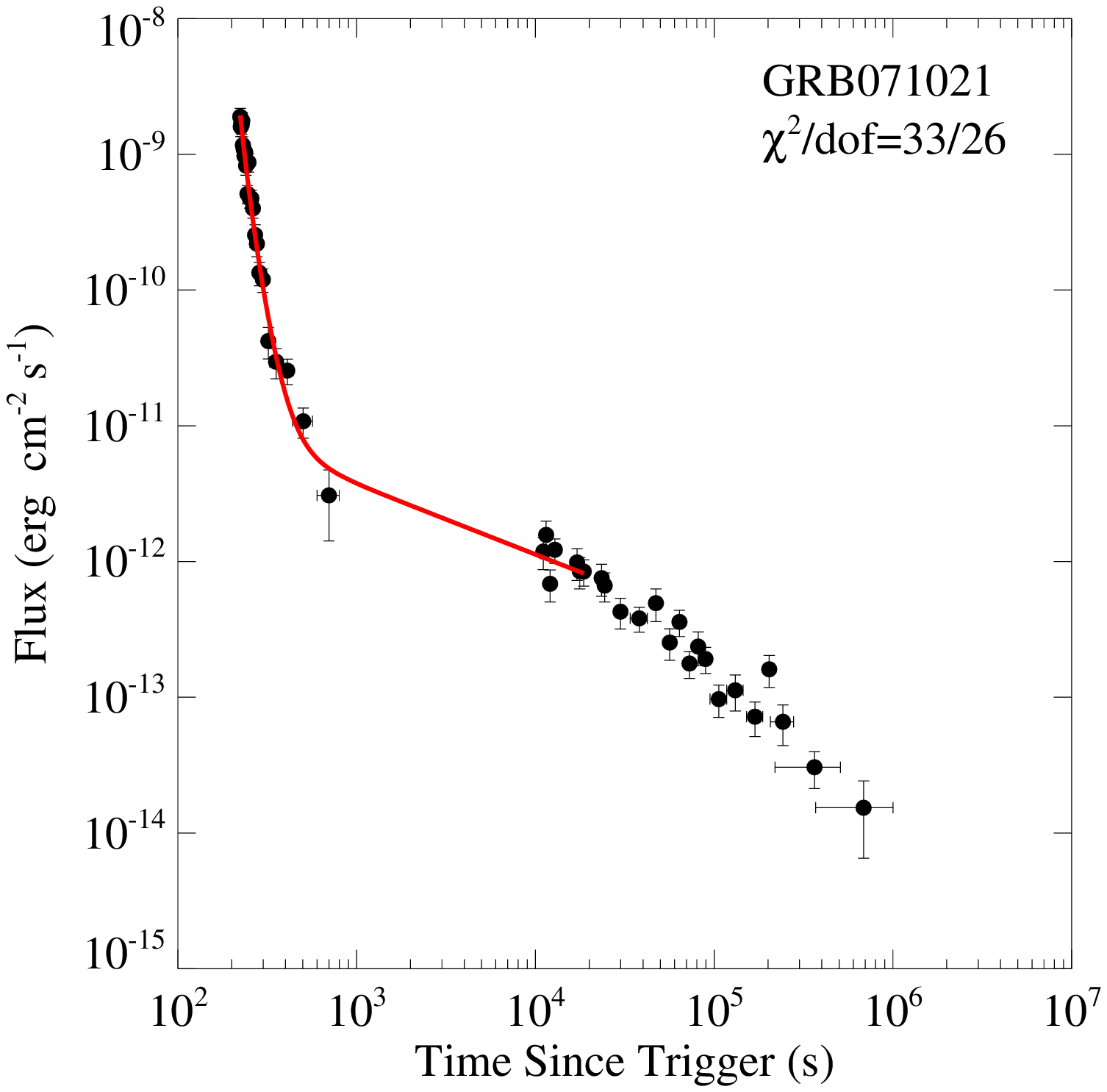}
\includegraphics[angle=0,scale=0.35]{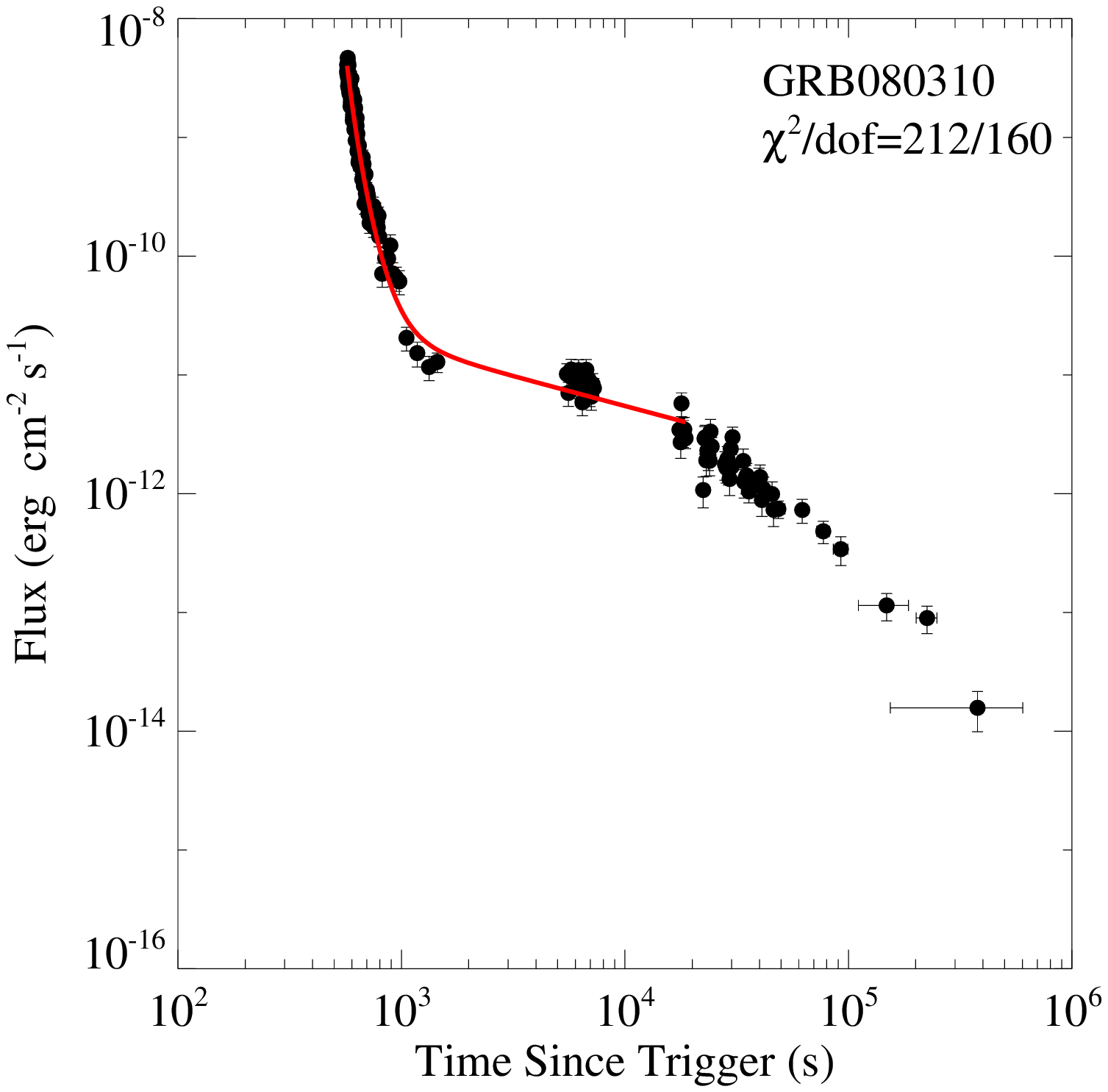}
\includegraphics[angle=0,scale=0.35]{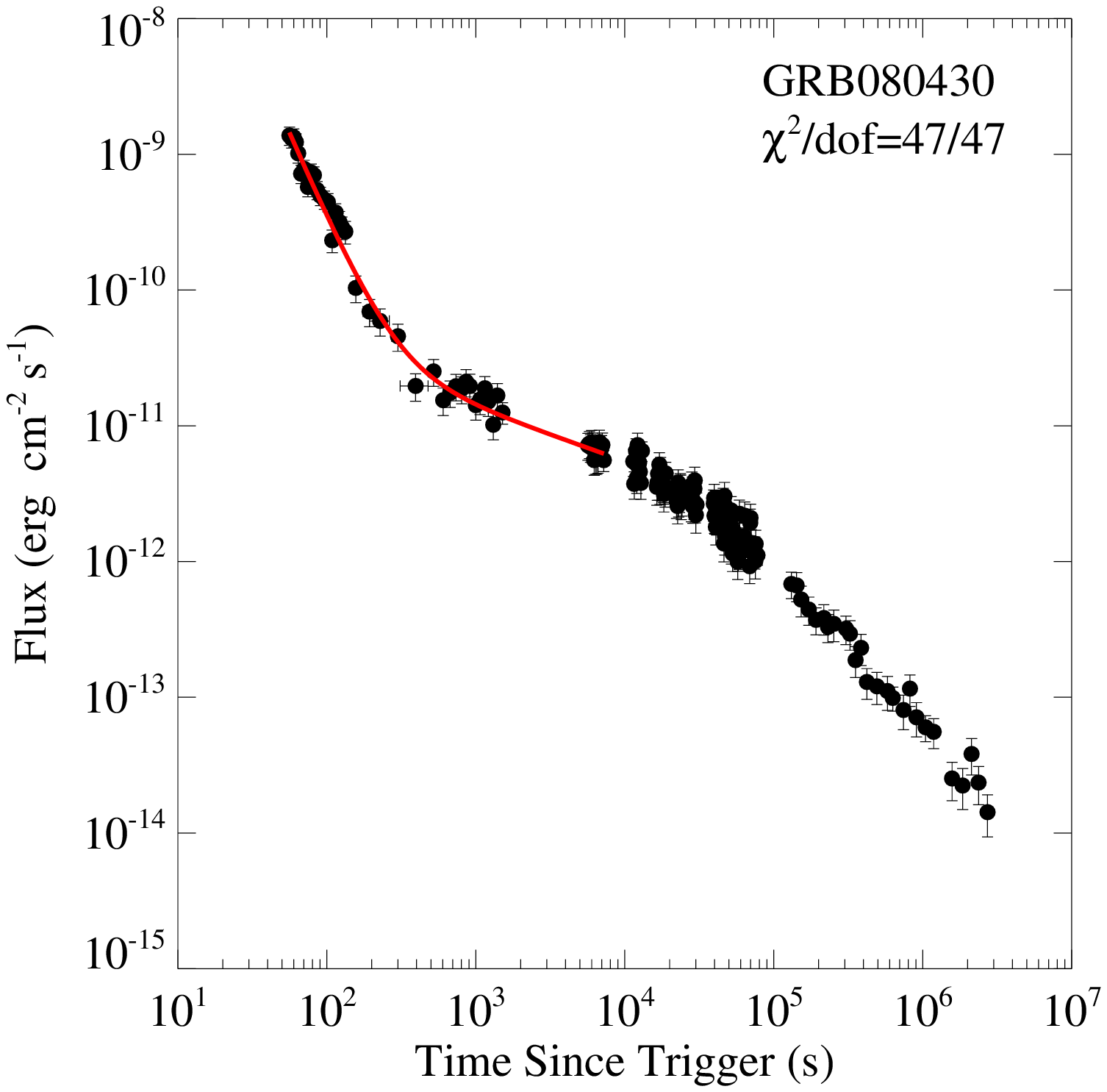}
\includegraphics[angle=0,scale=0.35]{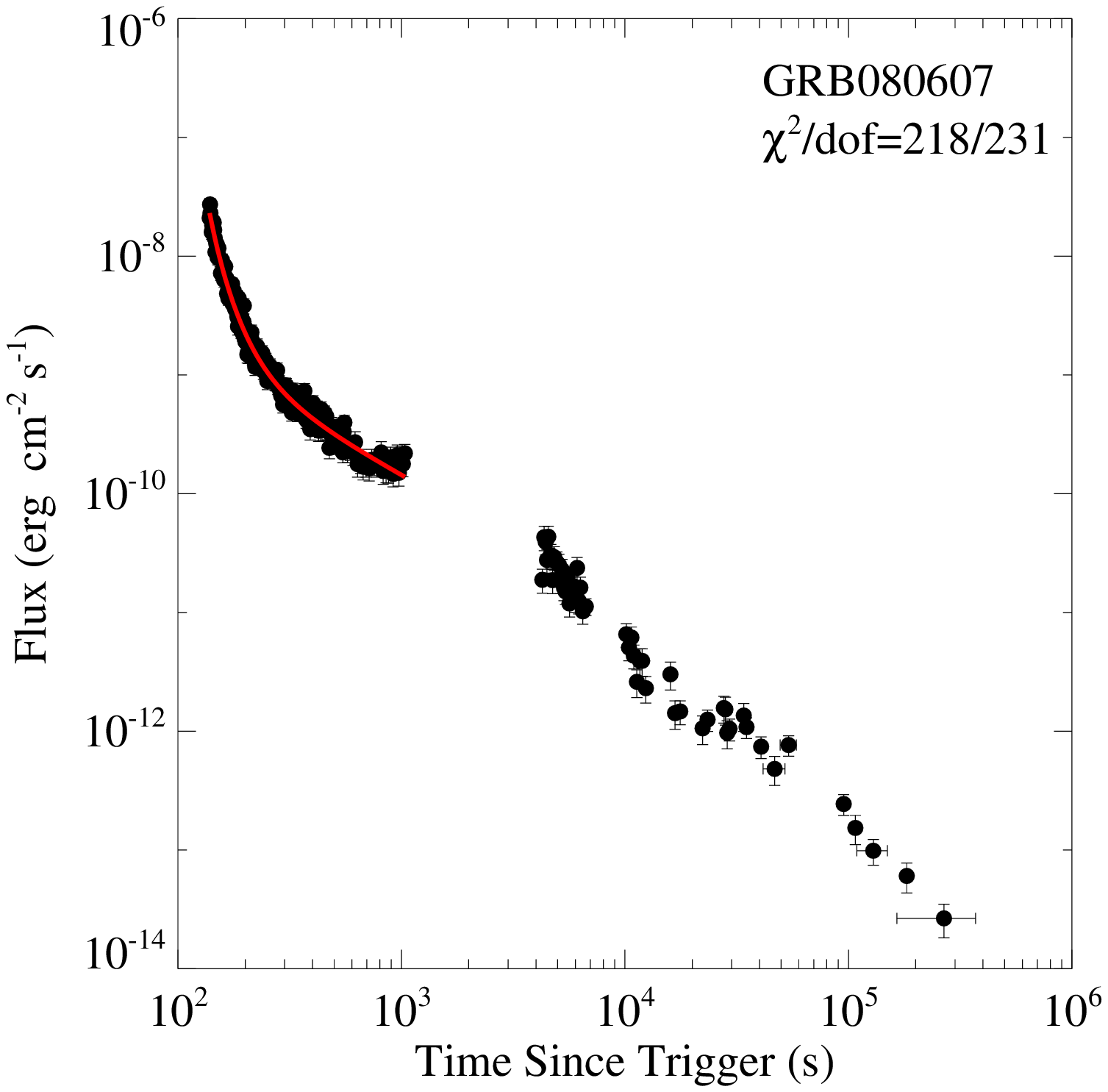}
\includegraphics[angle=0,scale=0.35]{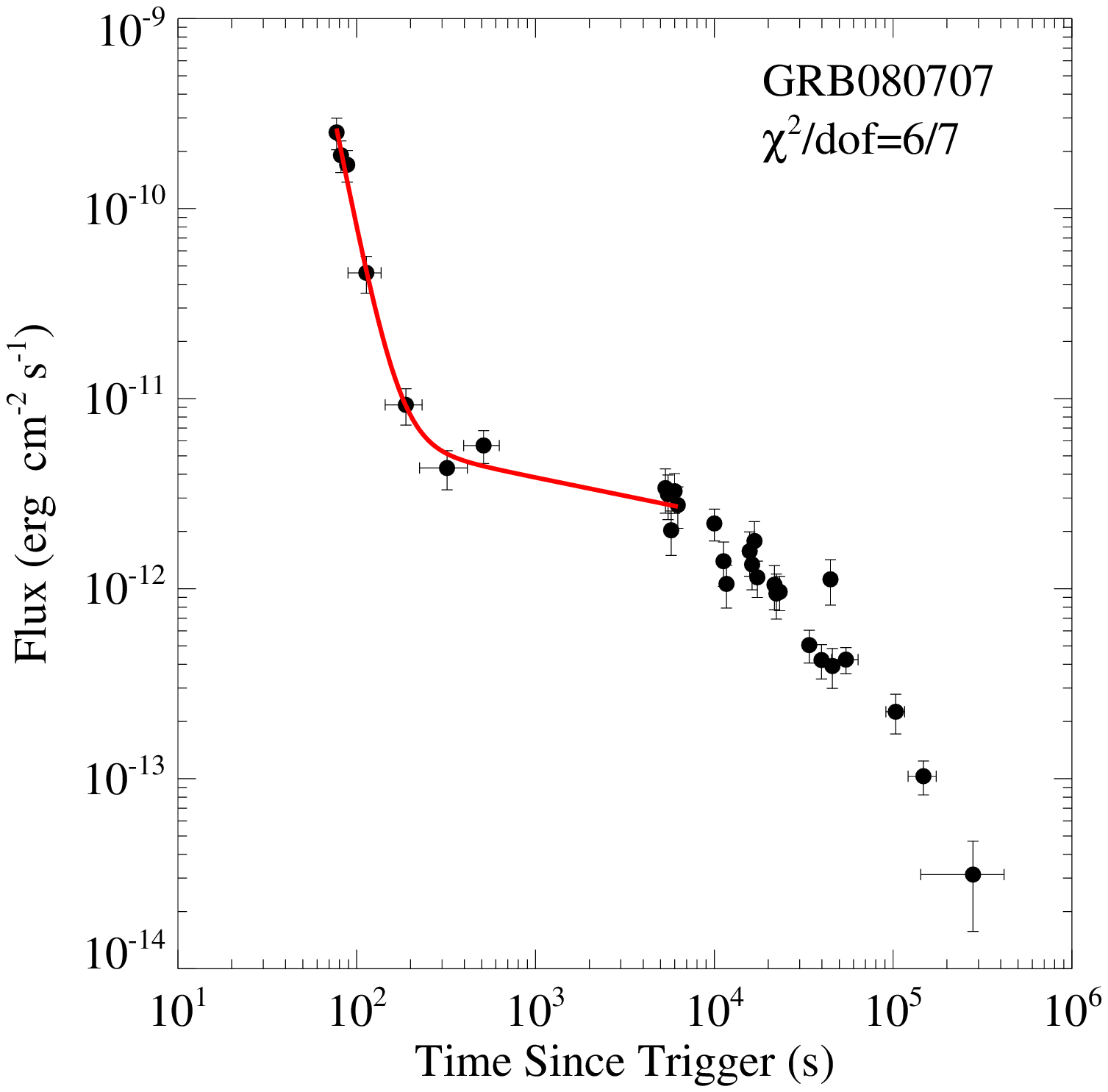}
\includegraphics[angle=0,scale=0.35]{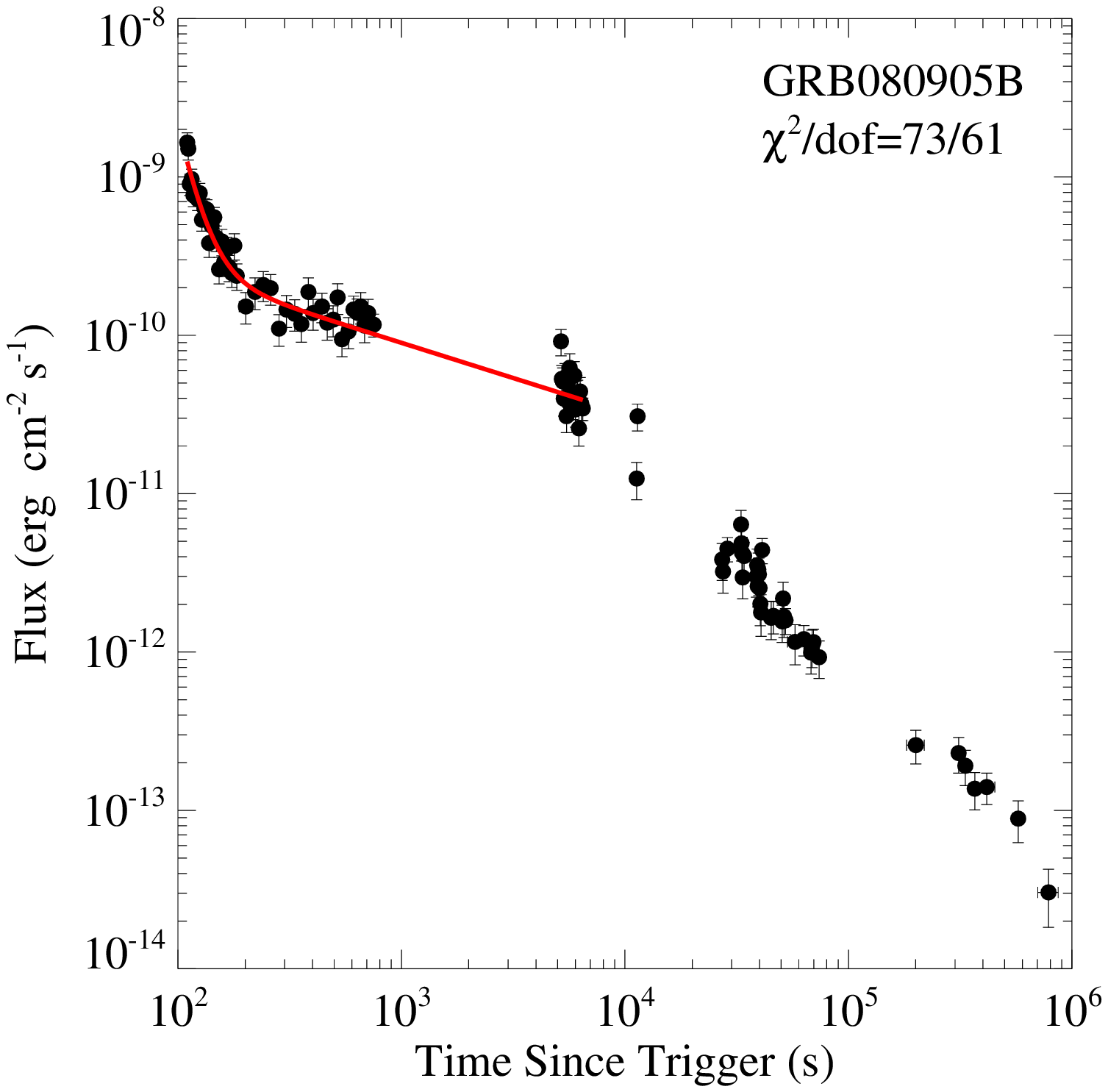}
\includegraphics[angle=0,scale=0.35]{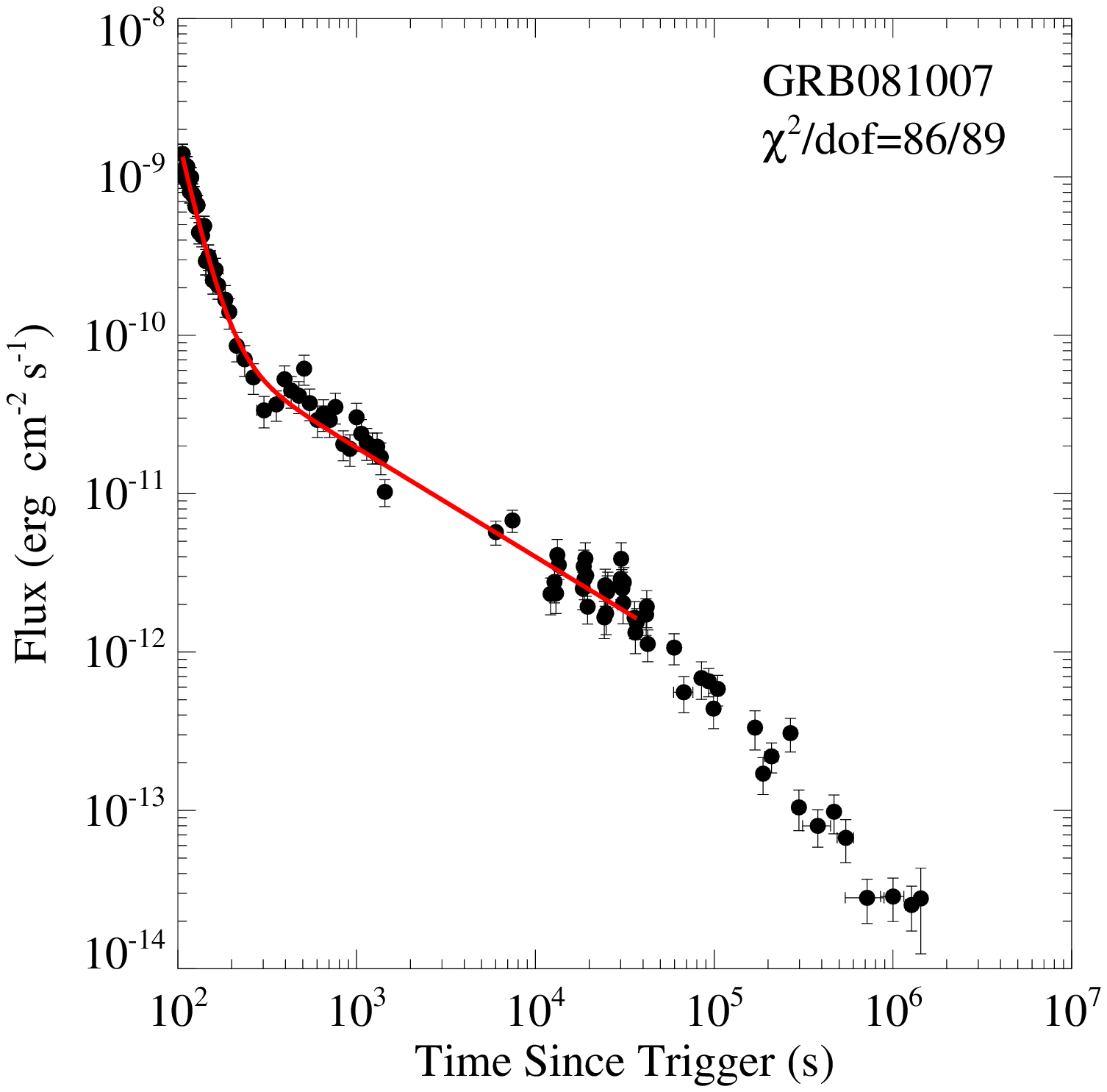}
\includegraphics[angle=0,scale=0.35]{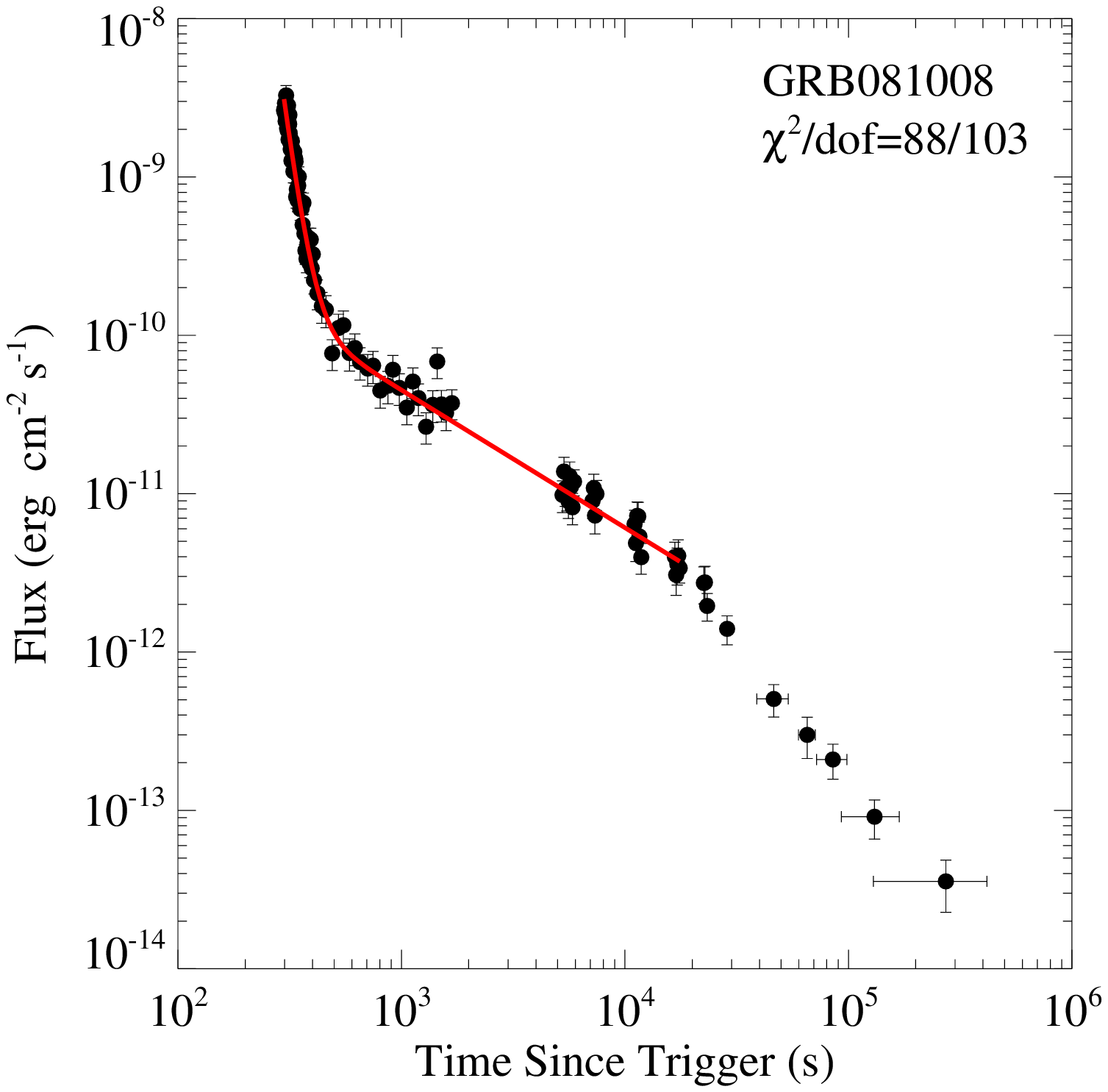}
\hfill \center{Fig.1  (continued)}
\end{figure*}

\begin{figure*}
\includegraphics[angle=0,scale=1]{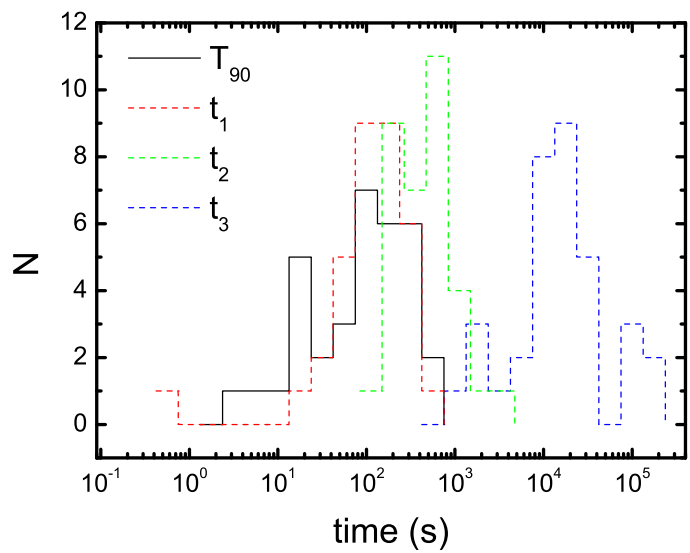}
\caption{Distributions of logarithmic $T_{90}$, $t_1$, $t_2$, and
$t_3$.}\label{cor}
\end{figure*}

\begin{figure*}
\includegraphics[angle=0,scale=0.72]{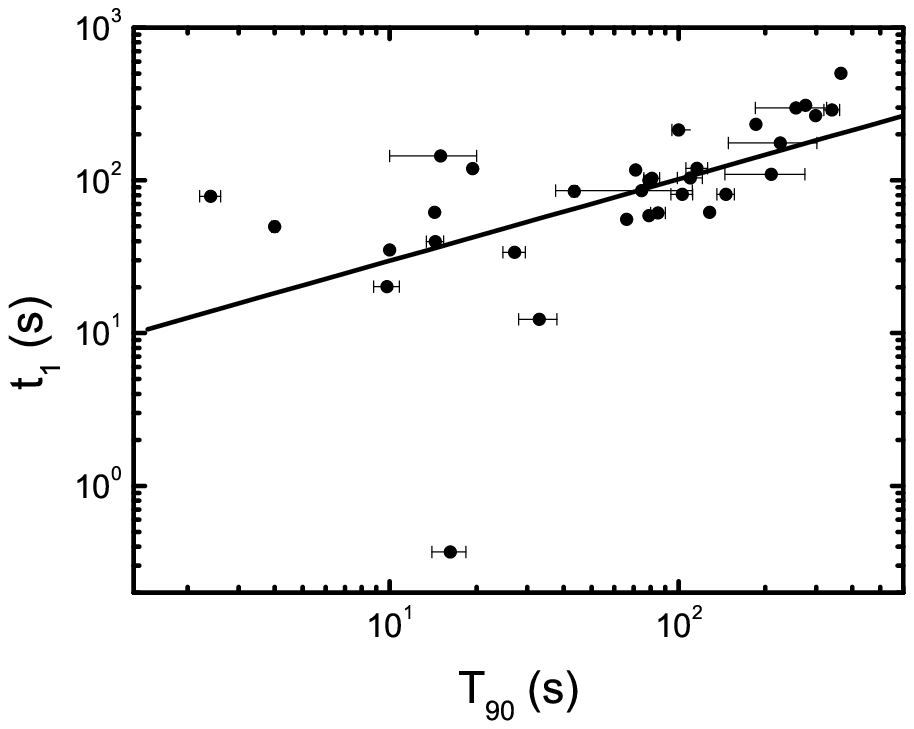}
\includegraphics[angle=0,scale=0.72]{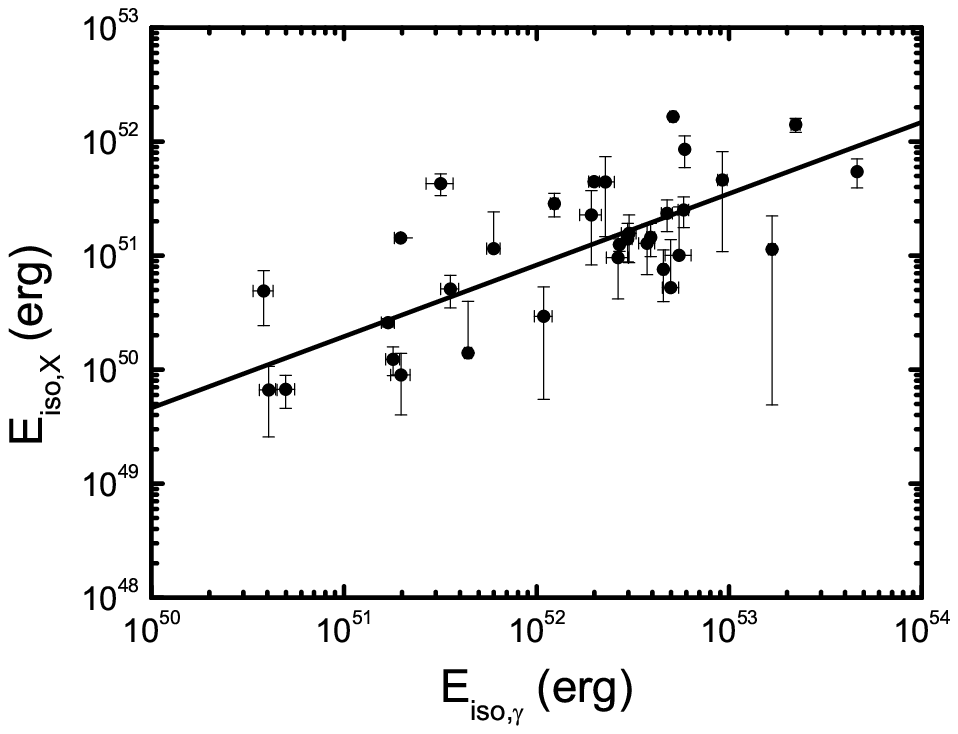}
\includegraphics[angle=0,scale=0.72]{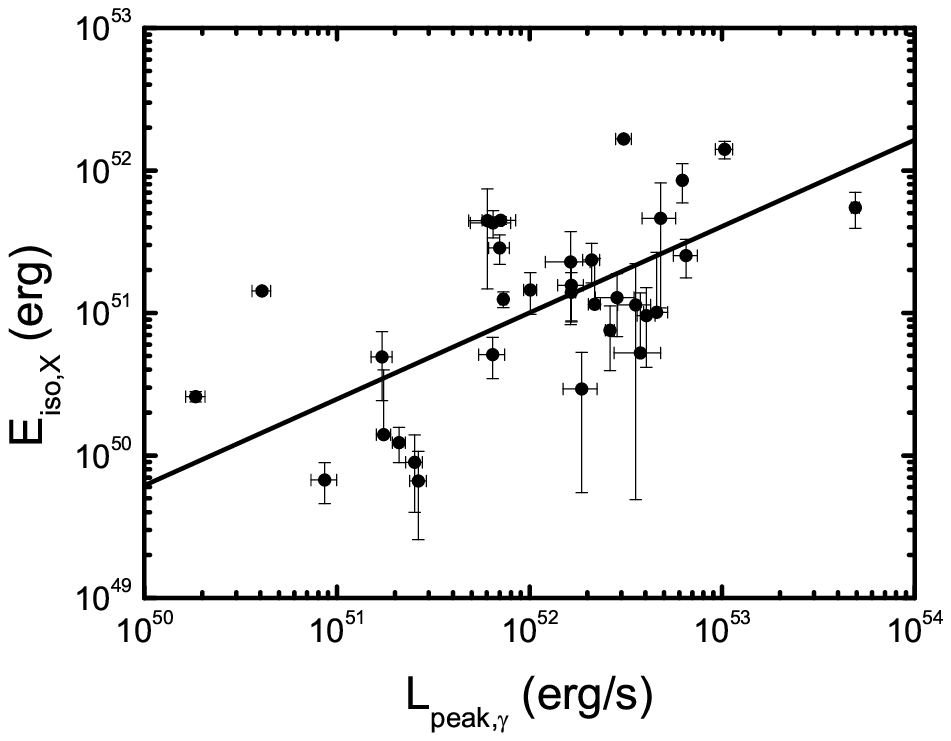}
\includegraphics[angle=0,scale=0.72]{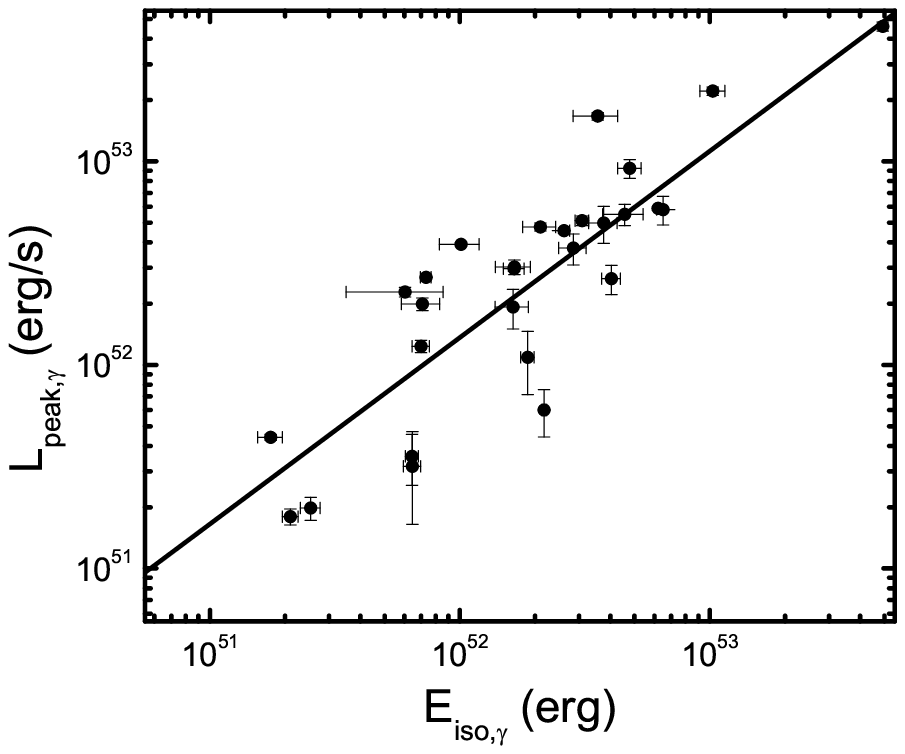}
\includegraphics[angle=0,scale=0.72]{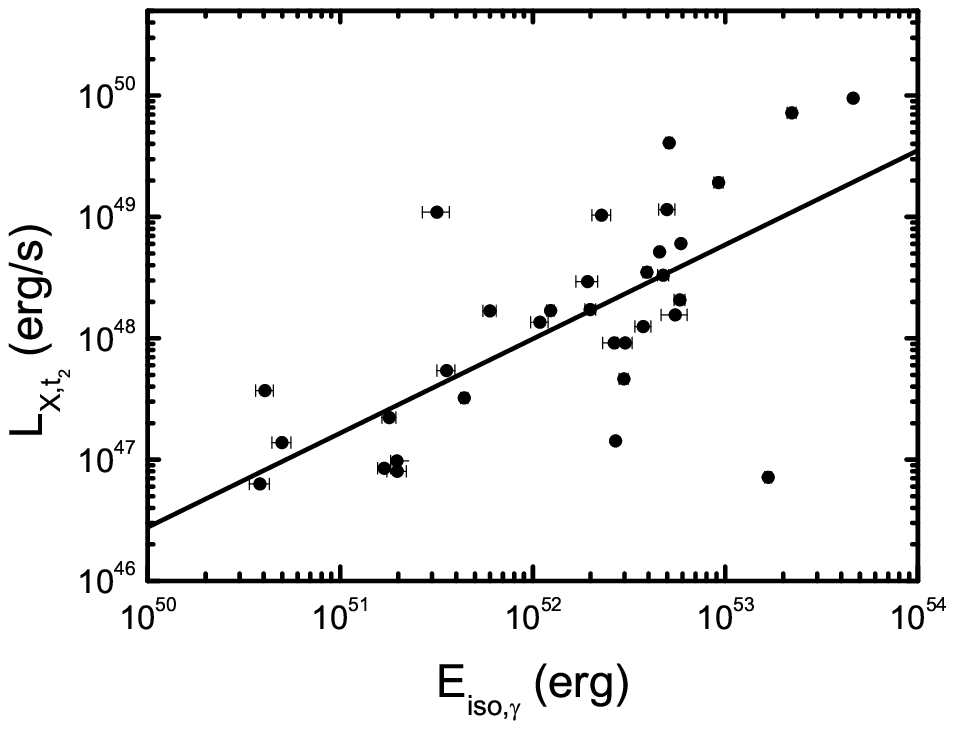}
\includegraphics[angle=0,scale=0.72]{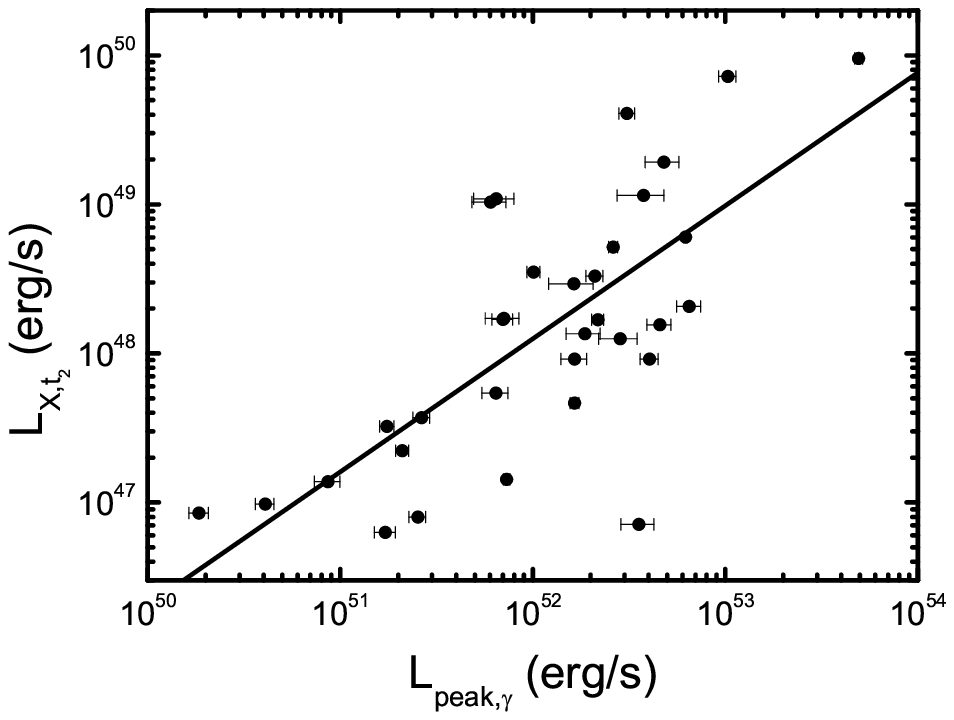}
\caption{Correlation between the properties of the prompt gamma-ray
phase and the shallow decay phase.  Lines are the best fitting
results.}\label{cor}
\end{figure*}

\begin{figure*}
\includegraphics[angle=0,scale=0.7]{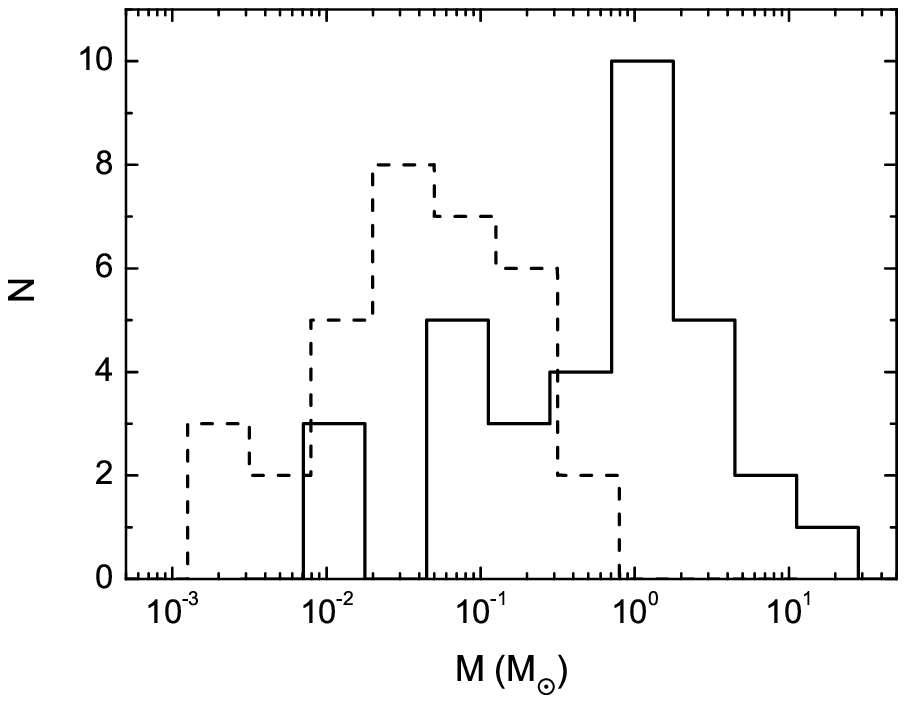}
\includegraphics[angle=0,scale=0.7]{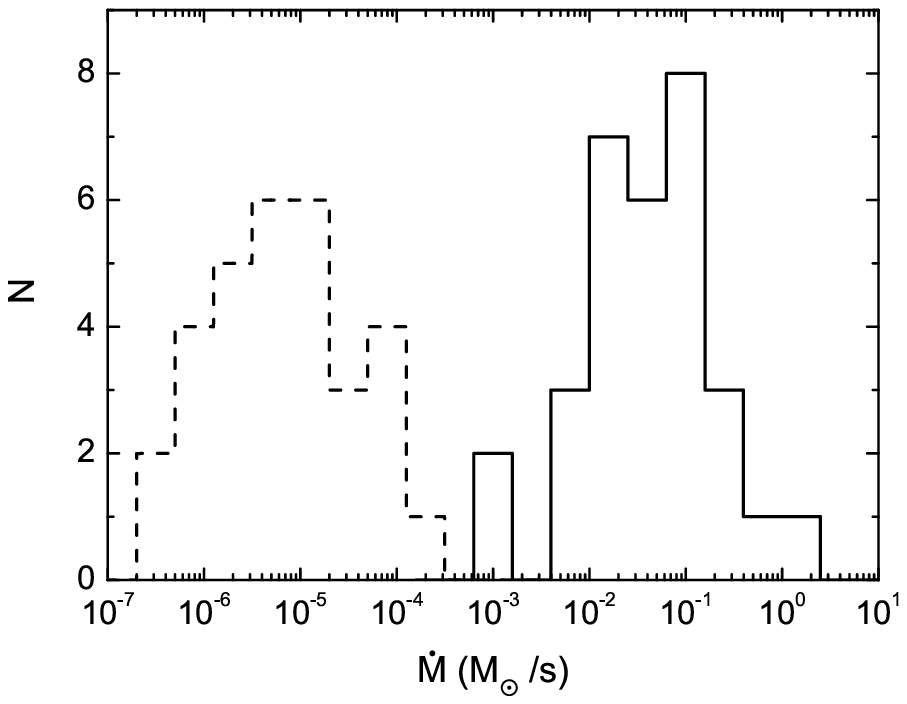}
\includegraphics[angle=0,scale=0.7]{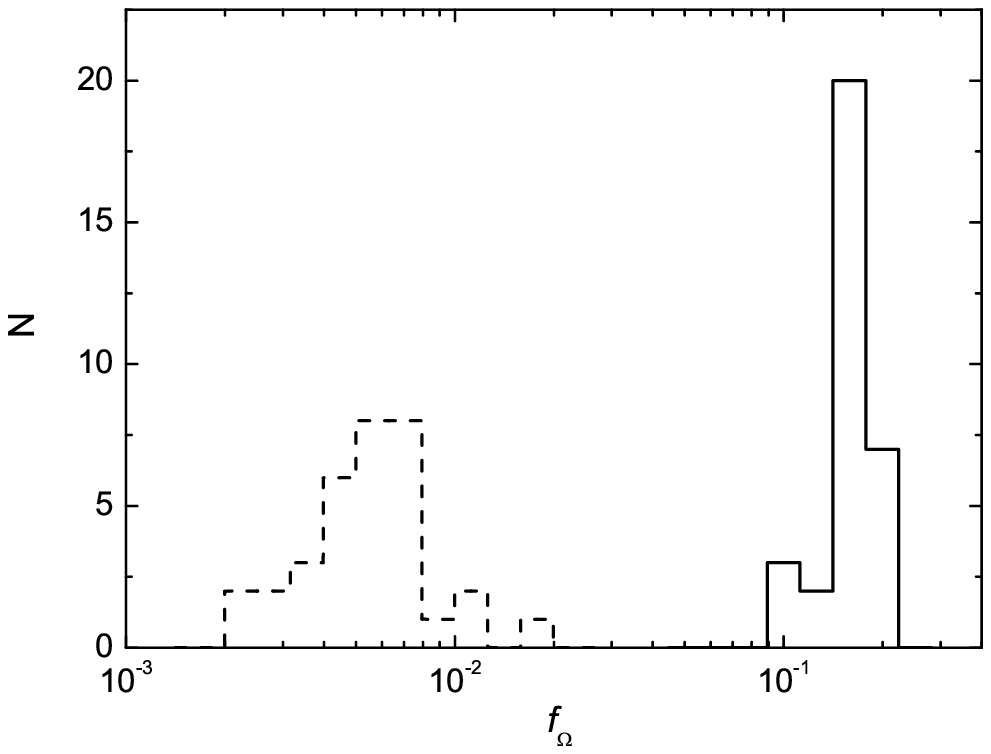}
%\hfill
\includegraphics[angle=0,scale=0.7]{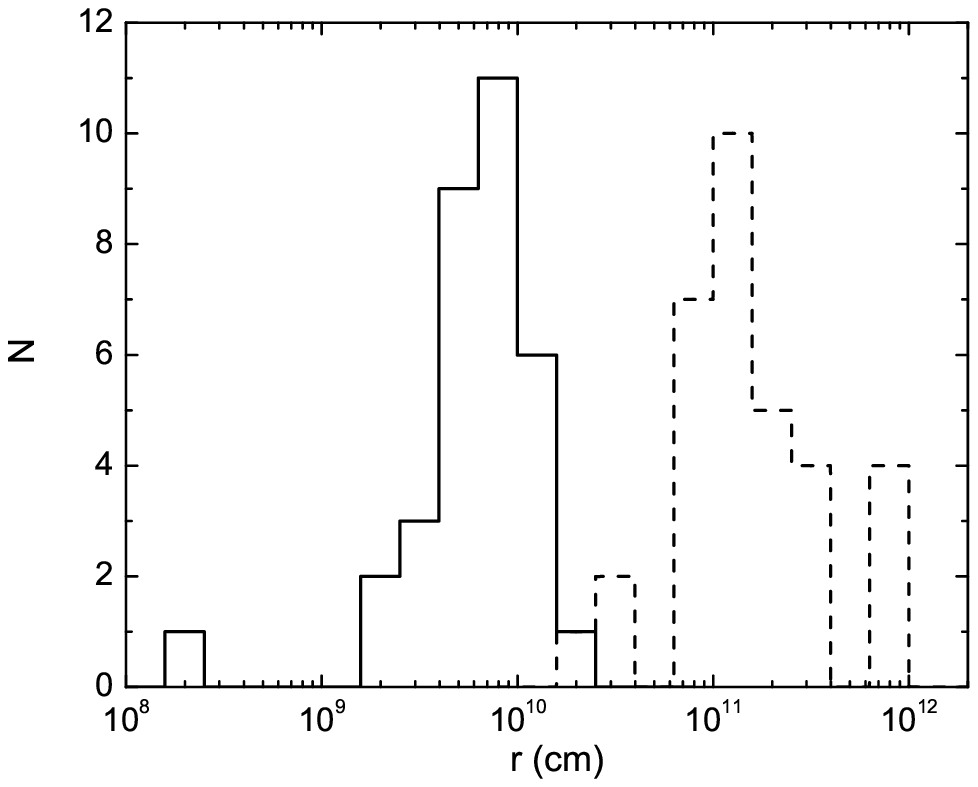}
\caption{Distributions of the derived properties of the core ({\em solid}) and envelope ({\em dashed})
layers.}\label{cor}
\end{figure*}

\begin{figure*}
\includegraphics[angle=0,scale=0.8]{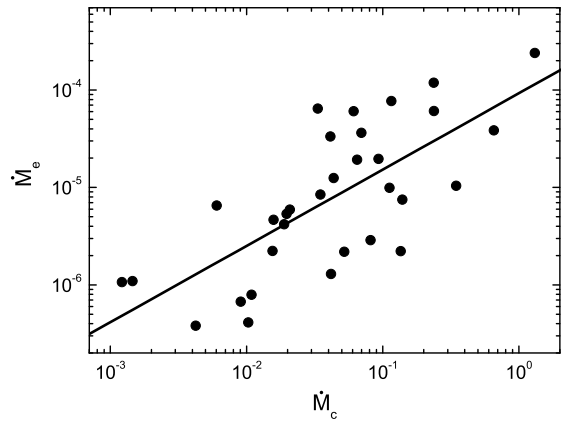}
\caption{Correlation of the average accretion rates between the
prompt gamma-rays $\dot{M}_{c}$ and the shallow decay phases
$\dot{M}_{e}$. The solid line is the best fit to the data with a
Spearman correlation coefficient $r=0.72$ and chance probability
$p<10^{-4}$.}\label{cor}
\end{figure*}

\begin{figure*}
\includegraphics[angle=0,scale=0.7]{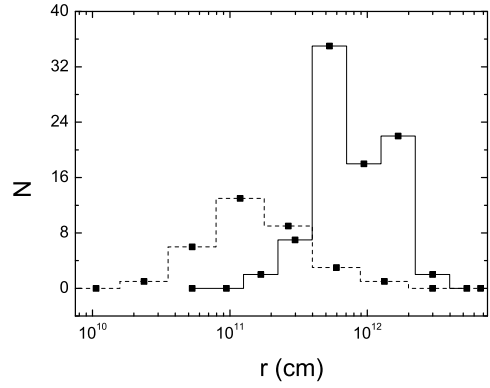}
\caption{Comparison of the radii of the envelope layers of the
bursts in our sample ({\em step dash line} ) with the photospherical
radii of 86 WR stars ({\em step solid line}) }\label{WN}
\end{figure*}

\begin{figure*}
\includegraphics[angle=0,scale=0.7]{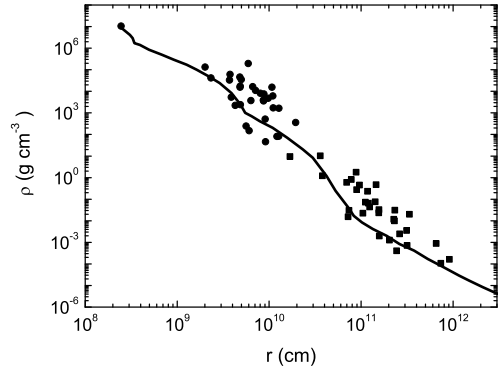}
\caption{Assembled mass density as a function of radius for the
bursts in our sample with comparison of the simulation for a
pre-supernova star with mass of 25$M_{\odot}$ (the
curve).}\label{cor}
\end{figure*}

\begin{figure*}
\includegraphics[angle=0,scale=0.7]{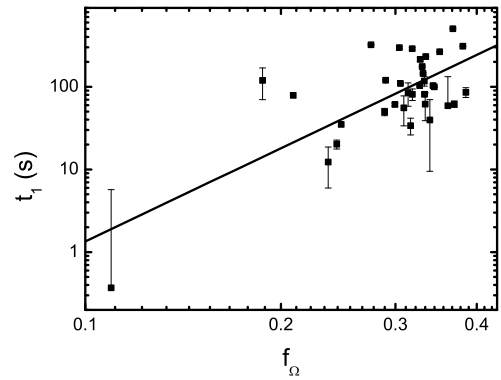}
\caption{Correlation between the rotational parameter and the burst
duration.}\label{cor}
\end{figure*}

\end{document}